\documentclass[final,3p,times]{elsarticle}

\journal{Neural Networks}

\usepackage{amsmath,amssymb,amsfonts}
\usepackage{algorithmic}
\usepackage{algorithm}
\usepackage{graphicx}
\usepackage{booktabs}
\usepackage{tablefootnote}
\usepackage{threeparttable}
\usepackage{subfigure}
\usepackage{enumitem}
\usepackage{xcolor}
\usepackage{xspace}
\usepackage{multirow}
\usepackage{array}
\usepackage{tabularx}
\usepackage{adjustbox}

\begin{document}

\begin{frontmatter}

\title{EMGFlow: Robust and Efficient Surface Electromyography Synthesis via Flow Matching}
\author[label1]{Boxuan Jiang}
\ead{data.j@sjtu.edu.cn}
\author[label1]{Chenyun Dai}
\ead{chenyundai@sjtu.edu.cn}
\author[label1]{Can Han\corref{corresponding}}
\ead{hancan@sjtu.edu.cn}

\cortext[corresponding]{Corresponding author.}

\affiliation[label1]{organization={School of Biomedical Engineering, Shanghai Jiao Tong University},
            city={Shanghai},
            postcode={200240}, 
            country={China}}

\begin{abstract}
Deep learning-based surface electromyography (sEMG) gesture recognition is frequently bottlenecked by data scarcity and limited subject diversity. 
While synthetic data generation via Generative Adversarial Networks (GANs) and diffusion models has emerged as a promising augmentation strategy, these approaches often face challenges regarding training stability or inference efficiency. 
To bridge this gap, we propose EMGFlow, a conditional sEMG generation framework. To the best of our knowledge, this is the first study to investigate the application of Flow Matching (FM) and continuous-time generative modeling in the sEMG domain. 
To validate EMGFlow across three benchmark sEMG datasets, we employ a unified evaluation protocol integrating feature-based fidelity, distributional geometry, and downstream utility. 
Extensive evaluations show that EMGFlow outperforms conventional augmentation and GAN baselines, and provides stronger standalone utility than the diffusion baselines considered here under the train-on-synthetic test-on-real (TSTR) protocol. 
Furthermore, by optimizing generation dynamics through advanced numerical solvers and targeted time sampling, EMGFlow achieves improved quality-efficiency trade-offs. 
Taken together, these results suggest that Flow Matching is a promising and efficient paradigm for addressing data bottlenecks in myoelectric control systems.
Our code is available at: https://github.com/Open-EXG/EMGFlow.
\end{abstract}



\begin{keyword}
Surface electromyography \sep Flow matching \sep Synthetic data generation \sep Deep generative models \sep Neural decoding
\end{keyword}

\end{frontmatter}

\section{Introduction}
\label{sec:introduction}
Surface electromyography  (sEMG)-based gesture recognition is an important enabling technology for human--machine interaction, rehabilitation, and prosthetic control~\cite{kaifosh2025generic}. Compared with vision-based or contact-rich sensing modalities, sEMG directly captures muscle activation associated with voluntary movement while remaining non-invasive, wearable, and suitable for continuous deployment in real-world settings~\cite{xu2024chatemg,sivakumar2024emg2qwerty}. These properties make sEMG a particularly attractive interface for intuitive control, where reliable decoding of user intent is essential for practical usability.

Recent advances in deep learning have substantially improved sEMG-based gesture recognition by enabling end-to-end modeling of multi-channel signals and by better capturing their complex spatiotemporal structure~\cite{yang2025stcnet,al2024tcnn,wang2024transformer}. Nevertheless, modern deep models remain vulnerable to overfitting in this domain because informative sEMG data are still limited: large-scale collection is labor-intensive and expensive, annotation is nontrivial, and the resulting datasets often exhibit substantial subject variability, recording noise, and protocol-dependent shifts. In addition, repeated trials within a session and overlapping sliding-window preprocessing can introduce considerable redundancy, so the effective diversity of training samples is often smaller than the nominal dataset size suggests. As a result, data scarcity and limited diversity continue to constrain the generalization ability of sEMG recognition systems~\cite{tsinganos2020data}.

Data augmentation therefore provides a practical approach to alleviating these limitations. Existing sEMG augmentation strategies can be broadly divided into two categories. The first consists of single-sample transformations, such as temporal perturbation, magnitude variation, resampling, frequency masking, and mix-based operations, which are simple to implement and often effective in practice~\cite{tsinganos2020data,zhao2024dominant,cao2024survey}. However, these methods act locally on existing windows and usually provide only limited variation around the observed data. The second category is generative augmentation, which attempts to learn the underlying data distribution and synthesize new samples directly. Prior studies have explored Generative Adversarial Networks (GANs)~\cite{mendez2022emg,ao2024overcoming,coelho2023novel}, transformer-based generative models~\cite{bird2021synthetic}, and, more recently, Denoising Diffusion Probabilistic Models (DDPMs)~\cite{ho2020denoising,dhariwal2021diffusion,xiong2024patchemg}. More broadly, recent reviews identify deep generative modeling as an increasingly active direction for physiological signals~\cite{neifar2025deepa}. Together, these works suggest that generative augmentation is a promising direction for sEMG, especially when the goal is to increase sample diversity beyond handcrafted local transformations.

At the same time, two important gaps remain. 
First, while GANs and diffusion models show promise, they are often hindered by inherent challenges: GANs are highly susceptible to training instability and mode collapse, whereas diffusion models typically require hundreds of sampling steps, severely limiting efficient generation.
To overcome these bottlenecks, Flow Matching (FM) and related continuous-time generative methods have emerged as a powerful generative paradigm because they directly learn continuous transport dynamics and naturally support flexible numerical solvers at inference time~\cite{liu2022flow,lipman2022flow}. Yet their role in sEMG generation and augmentation remains largely unexplored. Second, evaluation in this area is still incomplete. Many prior studies focus mainly on augmentation accuracy or a small set of fidelity-like metrics, while placing insufficient emphasis on the standalone utility of synthetic data, the coverage of the generated distribution, and the relationship between sample fidelity and downstream effectiveness. 
For sEMG, where the clinical value of synthetic data depends on their ability to support robust decoding of real human intent, evaluating standalone utility is not merely an alternative metric, but an important prerequisite for practical deployment.

To bridge these gaps, we propose EMGFlow, a conditional sEMG generation framework. To our knowledge, this is the first study to investigate continuous-time generative dynamics for sEMG synthesis, offering a practical alternative for addressing the data bottleneck in myoelectric control. To comprehensively validate our approach, we benchmark Flow Matching against conventional augmentation methods as well as representative generative baselines including GANs and DDPMs, under both augmentation and train-on-synthetic test-on-real (TSTR) settings. We further complement downstream evaluation with feature-based fidelity metrics and distributional diagnostics so that sample realism, coverage, and task utility can be examined jointly rather than in isolation. Beyond framework-level comparison, we also provide a systematic empirical analysis of design choices that are particularly relevant for Flow Matching in practice, including conditional interface design, normalization strategy, classifier-free guidance, time-sampling strategy, and solver choice.
These analyses reveal an important empirical finding in sEMG generation: high apparent class discriminability does not necessarily guarantee improved downstream performance. Furthermore, design choices during training and inference can substantially affect the trade-offs among fidelity, coverage, efficiency, and utility.

The main contributions of this work are summarized as follows:
\begin{enumerate}
    \item To the best of our knowledge, we are the first to investigate the application of Flow Matching and continuous-time generative modeling for sEMG synthesis. 
    Furthermore, we establish a unified evaluation paradigm that uniquely integrates standard augmentation, the train-on-synthetic test-on-real (TSTR) setting, feature-based fidelity, and distributional geometry.
    \item We show that Flow Matching outperforms conventional augmentation baselines and remains competitive with strong generative baselines such as GANs and DDPMs. Its advantage is particularly evident under the stricter TSTR setting, indicating stronger standalone utility of the generated synthetic sEMG data.
    \item We conduct an in-depth exploration of design choices in conditional Flow Matching for sEMG generation. Our findings show that appropriate conditioning and normalization design, balanced classifier-free guidance, advanced numerical solvers, and targeted time sampling strategies together yield improved quality-efficiency trade-offs, providing practical guidance for robust sEMG data synthesis.
\end{enumerate}

\begin{figure}[!t]
    \centering
    \includegraphics[width=0.92\textwidth]{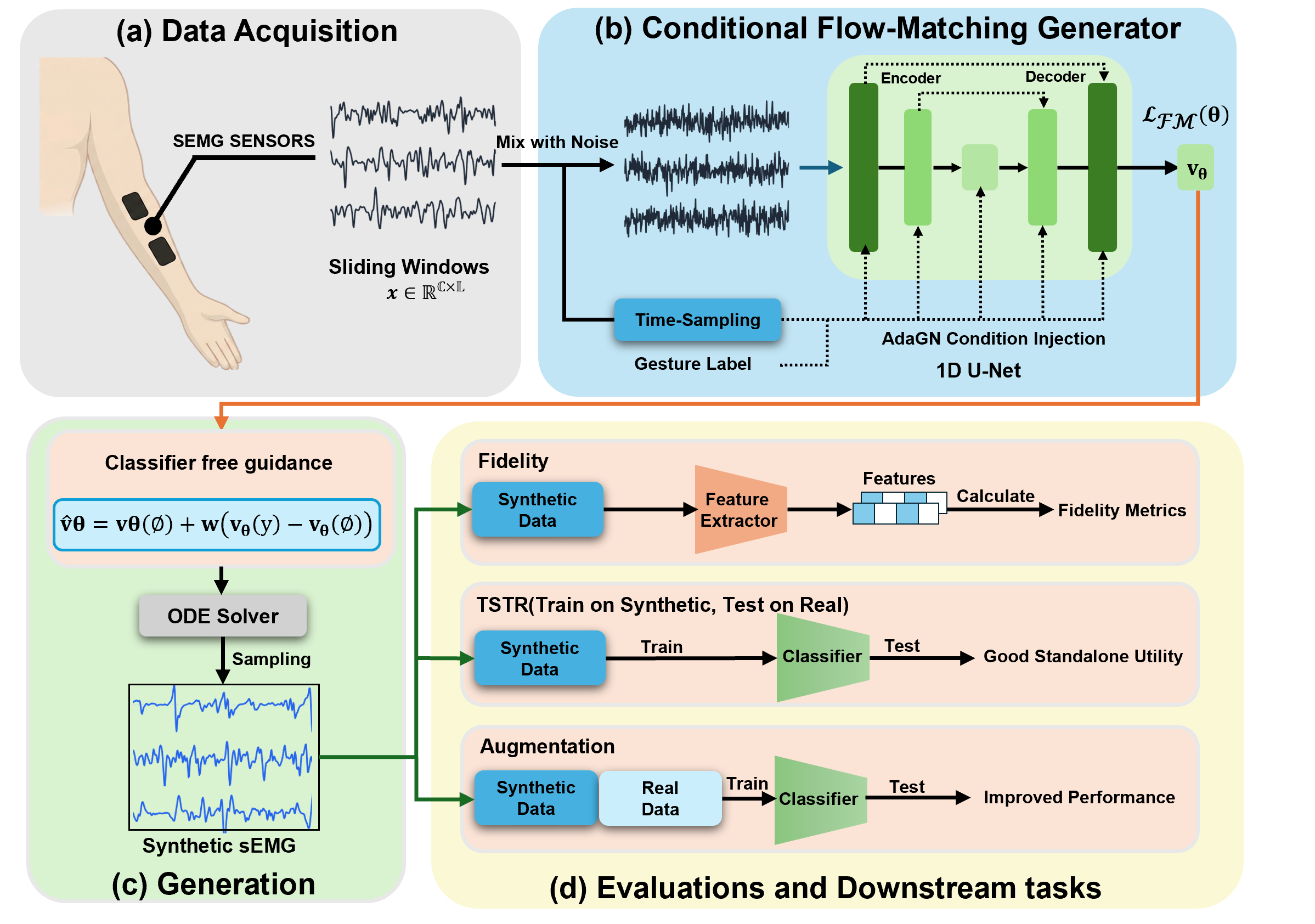}
    \caption{Overview of the proposed EMGFlow pipeline. The framework consists of four stages: (a) sEMG data acquisition and sliding-window preprocessing; (b) conditional flow-matching training with time sampling and AdaGN-based condition injection; (c) synthetic EMG generation via classifier-free guidance and ODE solvers; and (d) comprehensive evaluation through fidelity metrics, train-on-synthetic test-on-real (TSTR), and augmentation experiments.}
    \label{fig:emgflow_pipeline}
\end{figure}

\section{Related Works}

\subsection{Deep Learning for sEMG-based Gesture Recognition}
sEMG signals reflect muscle activation patterns and have been widely used for gesture recognition in human--machine interaction, prosthetic control, and related assistive applications~\cite{kaifosh2025generic,atzori2014characterization,meng2022usertailored}. Early studies mainly relied on handcrafted feature extraction and conventional classifiers, such as entropy- or decomposition-based features combined with SVM-style pipelines~\cite{prabhavathy2024hand}. More recently, deep learning has substantially improved sEMG-based gesture recognition by enabling end-to-end modeling of raw or minimally processed multi-channel signals. Representative efforts include improved cross-trial training schemes~\cite{dai2023improved}, hybrid convolutional-recurrent architectures such as EMGHandNet~\cite{karnam2022emghandnet}, compact convolutional models~\cite{al2024tcnn}, spatio-temporal cross networks such as STCNet~\cite{yang2025stcnet}, and transformer-based designs such as WaveFormer~\cite{chen2025waveformer}. These studies aim to better capture the complex, non-stationary, and multi-channel structure of sEMG signals so as to improve recognition accuracy and generalization. In contrast, our work focuses on synthetic data generation and augmentation as an orthogonal route to improving downstream performance.

\subsection{Data Augmentation for sEMG-based Gesture Recognition}

Data augmentation has long been regarded as a practical tool for improving generalization and mitigating overfitting in sEMG-based gesture recognition, especially when the available training data are limited~\cite{jiang2022optimization,tsinganos2020data}. Existing methods can be broadly divided into single-sample augmentation and generative augmentation. Single-sample methods apply label-preserving transformations directly to individual examples in the time, frequency, or mixed domains. Tsinganos \textit{et al.} provided one of the most systematic evaluations of sEMG augmentation strategies for hand gesture recognition, showing that properly designed signal transformations can yield substantial gains across benchmark datasets~\cite{tsinganos2020data}. Related augmentation ideas include temporal perturbation, masking, resampling, and mix-based operations~\cite{zhao2024dominant,cao2024survey,zhang2018mixup}. Similar principles have also been explored more broadly in wearable-sensor and time-series learning, including sensor-domain perturbations, frequency-domain augmentation, and diversity-oriented augmentation schemes~\cite{um2017data,semenoglou2023data,chen2023fraug,zhang2023towards}. Although these approaches are simple and computationally efficient, they usually generate only local variation around observed samples.

Generative augmentation attempts to address this limitation by learning the underlying data distribution and synthesizing new samples. Early efforts in biosignal generation include autoregressive language-model-style synthesis of EEG and EMG signals~\cite{bird2021synthetic}. In the sEMG domain, prior work has explored GAN-based augmentation for grasp or gesture recognition~\cite{mendez2022emg,coelho2023novel,ao2024overcoming}, diffusion-based few-shot EMG generation for data augmentation~\cite{xiong2024patchemg}, and autoregressive prompt-conditioned EMG generation for orthosis control in stroke rehabilitation~\cite{xu2024chatemg}. These studies show that learned generation can alleviate data scarcity and improve downstream robustness. However, the literature remains centered mainly on handcrafted augmentation, GAN-based generation, or task-specific generative designs, while modern continuous-time flow-based generative modeling has received little systematic study in sEMG gesture recognition. Our work addresses this gap by introducing Flow Matching as a modern generative baseline for conditional sEMG synthesis and by evaluating it under both augmentation and synthetic-training protocols.

\subsection{Generative Modeling for Physiological Signals and EMG Synthesis}

Generative modeling for physiological signals has become increasingly active in recent years, as summarized by recent reviews of deep generative models for biosignals~\cite{neifar2025deepa}. Within this broader space, generative approaches have been used for EEG, ECG, and EMG augmentation through autoregressive, adversarial, and diffusion-style models~\cite{bird2021synthetic,venugopal2024boosting}. For sEMG specifically, prior studies have investigated GAN-based augmentation for grasp classification, fatigue-robust gesture recognition, and WGAN-GP-based synthesis~\cite{mendez2022emg,ao2024overcoming,coelho2023novel}. More recently, diffusion-based methods have entered the field, with DDPMs providing a strong general-purpose generative baseline and PatchEMG adapting diffusion modeling to few-shot sEMG augmentation~\cite{ho2020denoising,dhariwal2021diffusion,xiong2024patchemg}. These studies collectively suggest that learned generative modeling is a promising direction for synthetic EMG generation.

However, two gaps remain. First, modern flow-based generative methods, despite their efficiency and flexibility in the broader generative modeling literature~\cite{lipman2022flow,liu2022flow}, have remained largely absent from sEMG generation. Second, evaluation in synthetic EMG studies is still often centered on augmentation accuracy or a narrow set of fidelity indicators, leaving standalone synthetic-data utility, support coverage, and distributional geometry insufficiently characterized. Our work addresses these gaps by introducing Flow Matching as a modern continuous-time generative baseline for sEMG and by establishing a more comprehensive evaluation framework that jointly assesses fidelity, coverage, and downstream utility.

\section{Methodology}

\subsection{Problem Setup}
\label{sec:problem_setup}

We consider conditional window-level sEMG generation for gesture recognition. For each subject, the multichannel sEMG stream is segmented into fixed-length windows, and each window is associated with a gesture label. Formally, a real sEMG sample is denoted as
\begin{equation}
x \in \mathbb{R}^{C \times L}, \qquad y \in \{1,\dots,K\},
\end{equation}
where $C$ is the number of EMG channels, $L$ is the window length, and $y$ is the gesture label.

Our goal is to learn a conditional generator
\begin{equation}
G(z,y) \rightarrow \hat{x} \in \mathbb{R}^{C \times L},
\end{equation}
which maps a random latent variable $z$ and a gesture label $y$ to a synthetic sEMG window $\hat{x}$ with realistic temporal patterns and class-consistent structure. The generated samples are used in two downstream settings: data augmentation and synthetic-only classifier training.

In the following, we instantiate this conditional generation problem using Flow Matching.

\subsection{Flow Matching for Conditional EMG Generation}
\label{sec:fm_conditional_emg}

We model conditional sEMG generation as learning a continuous-time transport from a tractable prior distribution to the class-conditional distribution of multichannel sEMG windows. Given a gesture label $y$, the objective is to generate a window-level sEMG sample whose temporal waveform and class-dependent structure are consistent with real data. To this end, we adopt Flow Matching, which learns a time-dependent velocity field that transports samples from noise to data along a prescribed probability path.

Let $x_1 \sim p_{\mathrm{data}}(x \mid y)$ denote a real multichannel sEMG window from class $y$, with $x_1 \in \mathbb{R}^{C \times L}$, and let $x_0 \sim p_0(x)$ denote a Gaussian noise sample of the same shape. Following the conditional Flow Matching formulation, we define a continuous interpolation path between $x_0$ and $x_1$ over time $t \in [0,1]$ by linear interpolation:
\begin{equation}
x_t = (1 - t)x_0 + t x_1.
\end{equation}
Under this path, the corresponding target velocity field is given by the time derivative of $x_t$:
\begin{equation}
\frac{d x_t}{d t} = x_1 - x_0.
\end{equation}

We then train a neural network $v_\theta(x_t, t, y)$ to predict this target velocity from the interpolated sample $x_t$, the time variable $t$, and the class label $y$. The resulting conditional Flow Matching objective is

\begin{equation}
\label{eq:fm_objective}
\mathcal{L}_{\mathrm{FM}}(\theta)
=
\mathbb{E}\!\left[
\left\|
v_\theta(x_t, t, y) - (x_1 - x_0)
\right\|_2^2
\right],
t \sim p(t).
\end{equation}

In this formulation, the network takes a noisy/interpolated multichannel EMG window $x_t$, a continuous time variable $t$, and a gesture label $y$ as input, and predicts a velocity field of the same shape as the EMG window. Here, $p(t)$ denotes the training-time sampling distribution over $t$. In the standard formulation, one may take $t \sim \mathcal{U}(0,1)$, while in our experiments we also investigate non-uniform $t$ sampling strategies and analyze their empirical effect on generation quality and downstream utility.

After training, generation is performed by solving the learned generative Ordinary Differential Equation (ODE) from $t=0$ to $t=1$, starting from an initial noise sample $x_0 \sim p_0(x)$:
\begin{equation}
\frac{d x_t}{d t} = v_\theta(x_t, t, y), \qquad x(0)=x_0.
\end{equation}

The synthetic sample is obtained as the terminal state $x(1)$, which is a generated multichannel sEMG window conditioned on the target gesture label. In practice, this ODE is solved numerically with a dedicated solver. Given a discretization $\{t_k\}_{k=0}^K$, the state is iteratively updated from $t_k$ to $t_{k+1}$ according to the chosen integration rule. For example, under the explicit Euler method,
\begin{equation}
x_{t_{k+1}} = x_{t_k} + (t_{k+1}-t_k)\, v_\theta(x_{t_k}, t_k, y).
\end{equation}

\subsection{Generative Baselines}
\label{sec:ddpm_shared_arch}

To benchmark Flow Matching against representative generative alternatives, we include a diffusion baseline trained with the standard DDPM objective~\cite{ho2020denoising} and a WGAN-GP baseline~\cite{coelho2023novel}. The diffusion model shares the same backbone architecture, input representation, and condition-injection mechanism as FM, while WGAN-GP is included as a representative GAN baseline, since adversarial models remain a strong baseline family in physiological signal generation~\cite{coelho2023novel,venugopal2024boosting}. At inference time, we evaluate the same trained diffusion model with two sampling configurations: 50-step DDIM sampling and full 1000-step ancestral DDPM sampling. To keep the comparison fair, we use the same lightweight U-Net backbone for FM and the diffusion baseline, and keep the WGAN-GP model size and overall training budget in the same regime, so that differences are not mainly driven by scale or optimization budget.

\subsection{Guidance, Time Sampling, and Solver Variants}
\label{sec:method_variants}

Beyond the generative framework itself, we study three factors that can substantially affect generation quality and downstream utility: classifier-free guidance, the training-time sampling distribution over $t$, and the numerical solver used at inference time.

\textbf{a) Classifier-free guidance:}
To control the strength of conditional generation at inference time, we adopt classifier-free guidance (CFG)~\cite{ho2021classifier}. During training, the generator is exposed to both conditional and unconditional inputs by randomly dropping the class condition with a fixed probability of 0.05. At sampling time, let $v_\theta(x_t, t, y)$ denote the conditional vector-field prediction and $v_\theta(x_t, t, \varnothing)$ denote the unconditional prediction, where $y$ is the class label and $\varnothing$ indicates a dropped condition. We then form the guided prediction as
\begin{equation}
\hat{v}_\theta(x_t, t, y)
=
v_\theta(x_t, t, \varnothing)
+
w \Bigl(
v_\theta(x_t, t, y) - v_\theta(x_t, t, \varnothing)
\Bigr),
\end{equation}
where $w \ge 1$ is the guidance weight. When $w=1$, the sampling process reduces to standard conditional generation without additional guidance amplification. Larger $w$ strengthens class-conditioning, but may also alter the coverage of the generated distribution.

In our experiments, CFG is applied only at inference time. Unless otherwise specified, the default guidance weight is set to $w=1.0$. To study its effect in the EMG generation setting, we further perform a systematic scan over $w \in \{1.0, 1.25, 1.5, 2.0, 2.5\}$ in the empirical analysis section.

\textbf{b) Time sampling:}
In Flow Matching, the distribution used to sample continuous time points $t$ acts as an important inductive bias during training, since it determines which parts of the trajectory receive more supervision. We compare uniform sampling against logit-normal time sampling in our experiments. Under uniform sampling,
\begin{equation}
t \sim \mathcal{U}(0,1).
\end{equation}
In contrast, logit-normal sampling draws
\begin{equation}
z \sim \mathcal{N}(\mu,\sigma^2), \qquad
t = \mathrm{sigmoid}(z)=\frac{1}{1+e^{-z}},
\end{equation}
which induces a non-uniform distribution on $t \in (0,1)$. Compared with uniform sampling, logit-normal sampling places relatively more probability mass on the middle portion of the trajectory and less on the extreme low-$t$ and high-$t$ regions. In our setting, this bias is beneficial, suggesting that emphasizing the intermediate regime can improve optimization and final sample quality.

\textbf{c) Solver variants:}
Since FM sampling is defined by integrating a learned ODE, the numerical solver can strongly affect sample quality under a fixed sampling budget. The generative trajectory is governed by
\begin{equation}
\frac{d x_t}{d t} = v_\theta(x_t,t,y),
\end{equation}
where $v_\theta$ is the learned velocity field. In general, explicit higher-order solvers update the state by combining multiple network evaluations within each step:
\begin{equation}
x_{n+1} = x_n + h \sum_{i=1}^{s} b_i\, v_\theta(\tilde{x}_i,\tilde{t}_i,y),
\end{equation}
where $h$ is the step size, and $s$ is the number of evaluations used in one step. Euler, Heun, and RK4 correspond to first-, second-, and fourth-order instances of this family, respectively. The key idea is that higher-order solvers query the learned velocity field multiple times within each step, which can reduce numerical error much faster than a single-evaluation update when the sampling budget is limited. This benefit, however, comes at the cost of increased function evaluations per step, so all comparisons are reported under matched numbers of function evaluations (NFE). For the diffusion baseline, we consider both accelerated DDIM sampling~\cite{song2020denoising} and full ancestral DDPM sampling in the main experiments, while the dedicated solver analysis focuses on DDIM as the practical accelerated counterpart.

\section{Experimental Protocol}
\subsection{Datasets and Preprocessing}

We evaluate the proposed framework on three public sEMG benchmarks from the Ninapro project, namely DB2, DB4, and DB7~\cite{atzori2014electromyography,pizzolato2017comparison,krasoulis2017improved}. These datasets are widely used for within-subject gesture recognition and cover different subject populations and gesture sets. Following common cross-trial evaluation protocols in recent sEMG recognition studies~\cite{dai2023improved,wang2024transformer}, we use trials 1, 3, 4, and 6 for training, and trials 2 and 5 for testing.

We uniformly preprocess all datasets, maintaining the raw 2000 Hz sampling rate without additional denoising. For each subject, the signals are segmented by a sliding window of 200 ms (400 samples) with a stride of 50 ms (100 samples). We then apply channel-wise z-score normalization within each subject, where the normalization statistics are computed from the training split and reused for the corresponding test split. Following common practice, the rest class is excluded and only gesture classes are retained for classification~\cite{dai2023improved,wang2024transformer}.

\begin{table}[!t]
    \caption{Characteristics and setup of three public sEMG datasets.} 
    \label{tab1}
    \centering
    \small
    \setlength{\tabcolsep}{4pt}
    \begin{tabular*}{\textwidth}{@{\extracolsep{\fill}}lccccccc@{}}
        \toprule
        Dataset & Subjects & Channels & Sampling (Hz) & Trials & Train & Test & Gestures \\
        \midrule
        Ninapro DB2 & 40 & 12 & 2000 & 6 & 1, 3, 4, 6 & 2, 5 & 49 \\
        Ninapro DB4 & 10 & 12 & 2000 & 6 & 1, 3, 4, 6 & 2, 5 & 52 \\
        Ninapro DB7 & 20 & 12 & 2000 & 6 & 1, 3, 4, 6 & 2, 5 & 40 \\
        \bottomrule
        \end{tabular*}
\end{table}

\subsection{Evaluation Settings}

We evaluate synthetic data under two complementary settings so as to assess both its practical usefulness and its standalone quality. The first is \textit{augmentation}, where synthetic samples are added to the real training set and the downstream classifier is then evaluated on the held-out real test set. This setting measures whether generated samples provide useful additional variation beyond the observed training data. The second is  the downstream classifier is then evaluated on the held-out real test set. This setting measures whether generated samples provide useful additional variation beyond the observed training data. The second is \textit{train-on-synthetic test-on-real} (TSTR), where the downstream classifier is trained using synthetic data and evaluated on the held-out real test set~\cite{esteban2017real}.

\subsection{Evaluation Metrics}

We evaluate synthetic data from three perspectives: downstream utility, sample fidelity, and distributional geometry. Unless otherwise specified, all feature-based metrics are computed in the latent space of a pretrained classifier trained on the real training split only.

For downstream utility, we report classification accuracy (ACC), macro-F1, and macro-recall under both augmentation and TSTR settings. For fidelity, we report Fr\'echet Inception Distance (FID)~\cite{heusel2017gans}, Inception Score (IS), and Category Accuracy Score (CAS), where CAS is defined as the accuracy of a classifier trained on the real training split and evaluated on generated samples. FID is additionally anchored against a real train-versus-test baseline. For distributional geometry, we report precision, recall, density, and coverage~\cite{kynkaanniemi2019improved,naeem2020reliable}, and additionally use neighborhood-based diagnostics such as KNN realism, train-test gap, and template concentration in the guidance analysis.

\subsection{Experimental Settings}

For both the diffusion baseline and Flow Matching (FM), we use the same compact 1D U-Net backbone adapted from PatchEMG~\cite{xiong2024patchemg}, while replacing the original BatchNorm-based design with GroupNorm and a stronger adaptive GroupNorm-style conditional modulation scheme. This serves as the default architecture unless otherwise specified. Both generators are trained for 20{,}000 steps using Adam with a learning rate of $5\times10^{-4}$ and a batch size of 128, and we maintain an exponential moving average (EMA) of model weights starting from step 6{,}000, with decay 0.9999 for FM and 0.999 for DDPM. For FM, the default training configuration uses logit-normal time sampling with $\mu=0$ and $\sigma=1$, cosine annealing, and 20-step Heun sampling. The diffusion model is trained with the standard DDPM objective and evaluated using both 50-step DDIM sampling and full 1000-step ancestral DDPM sampling, unless otherwise specified. We also tested cosine annealing for DDPM, but did not observe gains and in some cases found slight degradation, so the reported DDPM results use the standard fixed learning-rate setting. In all augmentation, TSTR, and fidelity evaluations, generated samples are drawn in a class-balanced manner, and the guidance weight is fixed to $w=1.0$ unless stated otherwise.

For feature-based fidelity evaluation, we train a separate EMGHandNet classifier~\cite{karnam2022emghandnet} on the real training split only for 75 epochs with label smoothing 0.05, a 5-epoch warmup, and cosine learning-rate decay to mitigate overfitting. For downstream evaluation, we use EMGHandNet and WaveFormer~\cite{chen2025waveformer}, both trained with AdamW, learning rate $1\times10^{-3}$, weight decay $3\times10^{-4}$, batch size 256, and 100 epochs.

Unless otherwise specified, all reported results are first computed separately for each subject and then averaged across subjects. The reported mean and standard deviation, therefore, reflect subject-level performance rather than pooled window-level statistics.

\begin{table*}[!t]
\centering
\caption{Performance comparison of conventional and generative augmentation methods on Ninapro DB7 using EMGHandNet and Waveformer. Results are reported as mean $\pm$ std (\%). Best results are highlighted in \textbf{bold}; second-best are \underline{underlined}. Overall, learned generative augmentation is consistently stronger than most hand-crafted perturbation baselines, and EMGFlow achieves the best or tied-best performance in most metrics, indicating a favorable accuracy--efficiency trade-off relative to DDIM, DDPM, and WGAN-GP.}
\label{tab:db7_aug}
\begin{tabular}{lcccccc}
\toprule
\multirow{2}{*}{Methods} 
& \multicolumn{3}{c}{EMGHandNet} 
& \multicolumn{3}{c}{WaveFormer} \\
\cmidrule(lr){2-4} \cmidrule(lr){5-7}
& Acc & Macro-F1 & Macro-Rec 
& Acc & Macro-F1 & Macro-Rec \\
\midrule

Baseline 
& 73.05 $\pm$ 4.48 & 73.08 $\pm$ 4.36 & 73.38 $\pm$ 4.37
& 76.36 $\pm$ 3.81 & 76.39 $\pm$ 3.67 & 76.62 $\pm$ 3.63 \\

Replicate 
& 74.84 $\pm$ 4.13 & 74.95 $\pm$ 3.97 & 75.21 $\pm$ 3.98
& 77.08 $\pm$ 3.84 & 77.14 $\pm$ 3.60 & 77.37 $\pm$ 3.60 \\

\midrule
Jitter\&Scale 
& 73.37 $\pm$ 4.43 & 73.46 $\pm$ 4.11 & 73.64 $\pm$ 4.15
& 74.24 $\pm$ 4.78 & 74.32 $\pm$ 4.31 & 74.66 $\pm$ 4.34 \\

Upsample 
& 74.60 $\pm$ 3.81 & 74.63 $\pm$ 3.55 & 74.94 $\pm$ 3.55
& 78.55 $\pm$ 3.68 & 78.64 $\pm$ 3.44 & 78.96 $\pm$ 3.44 \\

Freq-Mask 
& 74.42 $\pm$ 5.06 & 74.51 $\pm$ 4.74 & 74.70 $\pm$ 4.75
& 76.17 $\pm$ 3.85 & 76.22 $\pm$ 3.54 & 76.53 $\pm$ 3.48 \\

Mixup 
& 75.66 $\pm$ 4.66 & 75.68 $\pm$ 4.53 & 75.92 $\pm$ 4.46
& 78.75 $\pm$ 4.42 & 78.83 $\pm$ 4.00 & 79.14 $\pm$ 4.00 \\

STAug
& 73.59 $\pm$ 3.80 & 73.54 $\pm$ 3.70 & 73.85 $\pm$ 3.74
& 76.51 $\pm$ 3.57 & 76.53 $\pm$ 3.33 & 76.77 $\pm$ 3.34 \\

Freq-Mix 
& 74.77 $\pm$ 3.71 & 74.81 $\pm$ 3.30 & 75.05 $\pm$ 3.32
& 76.82 $\pm$ 3.79 & 76.98 $\pm$ 3.50 & 77.23 $\pm$ 3.46 \\

\midrule

WGAN-GP~\cite{coelho2023novel} 
& 76.04 $\pm$ 3.74 & 76.17 $\pm$ 3.50 & 76.29 $\pm$ 3.57
& 78.76 $\pm$ 3.83 & 78.84 $\pm$ 3.51 & 79.09 $\pm$ 3.52 \\

DDPM~\cite{ho2020denoising}
& \underline{77.73 $\pm$ 4.24} & \underline{77.86 $\pm$ 3.80} & \underline{78.16 $\pm$ 3.84}
& \underline{79.77 $\pm$ 4.13} & \underline{79.84 $\pm$ 3.83} & \textbf{80.22 $\pm$ 3.79} \\

PatchEMG~\cite{xiong2024patchemg}
& 73.78 $\pm$ 3.87 & 73.93 $\pm$ 3.74 & 74.20 $\pm$ 3.66
& 78.17 $\pm$ 3.93 & 78.18 $\pm$ 3.68 & 78.47 $\pm$ 3.68 \\

DDIM~\cite{song2020denoising} 
& 77.41 $\pm$ 3.76 & 77.69 $\pm$ 3.44 & 77.82 $\pm$ 3.41
& 79.66 $\pm$ 4.34 & 79.82 $\pm$ 3.97 & \underline{80.10 $\pm$ 3.99} \\

EMGFlow 
& \textbf{78.26 $\pm$ 3.95} & \textbf{78.49 $\pm$ 3.69} & \textbf{78.68 $\pm$ 3.71}
& \textbf{79.78 $\pm$ 3.97} & \textbf{79.90 $\pm$ 3.68} & \textbf{80.22 $\pm$ 3.63} \\

\bottomrule
\end{tabular}
\end{table*}

\begin{table*}[!t]
\centering
\caption{Performance comparison of conventional and generative augmentation methods on Ninapro DB4 using EMGHandNet and Waveformer. Results are reported as mean $\pm$ std (\%). Best results are highlighted in \textbf{bold}; second-best are \underline{underlined}. Similar to DB7, learned generative augmentation generally outperforms most conventional perturbation baselines, while EMGFlow provides the strongest overall performance among practical generators and remains competitive even against full-step DDPM.}
\label{tab:db4_aug}
\begin{tabular}{lcccccc}
\toprule
\multirow{2}{*}{Methods} 
& \multicolumn{3}{c}{EMGHandNet} 
& \multicolumn{3}{c}{Waveformer} \\
\cmidrule(lr){2-4} \cmidrule(lr){5-7}
& Acc & Macro-F1 & Macro-Rec 
& Acc & Macro-F1 & Macro-Rec \\
\midrule

Baseline 
& 64.74 $\pm$ 5.38 & 65.39 $\pm$ 5.41 & 65.65 $\pm$ 5.60
& 68.72 $\pm$ 5.76 & 69.17 $\pm$ 5.64 & 69.54 $\pm$ 5.75 \\

Replicate 
& 66.38 $\pm$ 5.48 & 67.05 $\pm$ 5.42 & 67.36 $\pm$ 5.62
& 68.50 $\pm$ 6.41 & 69.02 $\pm$ 6.28 & 69.40 $\pm$ 6.40 \\

\midrule
Jitter\&Scale 
& 65.35 $\pm$ 4.88 & 65.94 $\pm$ 4.74 & 66.26 $\pm$ 4.88
& 65.43 $\pm$ 5.72 & 66.08 $\pm$ 5.49 & 66.35 $\pm$ 5.62 \\

Upsample 
& 65.17 $\pm$ 5.18 & 65.79 $\pm$ 5.19 & 66.15 $\pm$ 5.31
& 69.02 $\pm$ 5.87 & 69.47 $\pm$ 5.78 & 69.86 $\pm$ 5.90 \\

Freq-Mask 
& 66.89 $\pm$ 4.84 & 67.39 $\pm$ 4.62 & 67.82 $\pm$ 4.90
& 67.52 $\pm$ 5.49 & 67.98 $\pm$ 5.48 & 68.26 $\pm$ 5.54 \\

Mixup 
& 68.24 $\pm$ 5.60 & 68.93 $\pm$ 5.68 & 69.15 $\pm$ 5.76
& 70.33 $\pm$ 6.04 & 70.95 $\pm$ 6.03 & 71.26 $\pm$ 6.07 \\

STAug
& 65.03 $\pm$ 4.79 & 65.68 $\pm$ 4.92 & 66.05 $\pm$ 4.97
& 67.06 $\pm$ 5.76 & 67.63 $\pm$ 5.69 & 67.98 $\pm$ 5.81 \\

Freq-Mix 
& 66.06 $\pm$ 5.54 & 66.80 $\pm$ 5.64 & 67.02 $\pm$ 5.69
& 68.11 $\pm$ 6.05 & 68.73 $\pm$ 5.98 & 69.03 $\pm$ 6.07 \\

\midrule

WGAN-GP~\cite{coelho2023novel} 
& 67.11 $\pm$ 4.71 & 67.74 $\pm$ 4.79 & 68.08 $\pm$ 4.96
& 69.94 $\pm$ 5.68 & 70.40 $\pm$ 5.62 & 70.82 $\pm$ 5.70 \\

DDPM~\cite{ho2020denoising}
& 69.94 $\pm$ 5.12 & 70.55 $\pm$ 4.98 & 70.85 $\pm$ 5.10
& 70.96 $\pm$ 5.69 & 71.45 $\pm$ 5.60 & 71.94 $\pm$ 5.67 \\

PatchEMG~\cite{xiong2024patchemg}
& 66.72 $\pm$ 5.22 & 67.24 $\pm$ 5.08 & 67.64 $\pm$ 5.28
& 70.17 $\pm$ 5.67 & 70.68 $\pm$ 5.59 & 71.09 $\pm$ 5.69 \\

DDIM~\cite{song2020denoising} 
& \underline{70.10 $\pm$ 4.74} & \underline{70.85 $\pm$ 4.68} & \underline{70.92 $\pm$ 4.83}
& \underline{71.27 $\pm$ 5.62} & \textbf{71.87 $\pm$ 5.57} & \underline{72.19 $\pm$ 5.65} \\

EMGFlow 
& \textbf{70.44 $\pm$ 4.74} & \textbf{71.10 $\pm$ 4.55} & \textbf{71.31 $\pm$ 4.73}
& \textbf{71.29 $\pm$ 5.80} & \underline{71.86 $\pm$ 5.72} & \textbf{72.20 $\pm$ 5.85} \\

\bottomrule
\end{tabular}
\end{table*}

\begin{table*}[!htbp]
\centering
\caption{Performance comparison of conventional and generative augmentation methods on Ninapro DB2 using EMGHandNet and Waveformer. Results are reported as mean $\pm$ std (\%). Best results are highlighted in \textbf{bold}; second-best are \underline{underlined}. EMGFlow remains among the strongest methods across both backbones, although full-step DDPM is slightly better on selected EMGHandNet metrics, suggesting that its small accuracy advantage comes at a substantially higher sampling cost.}
\label{tab:db2_aug}
\begin{tabular}{lcccccc}
\toprule
\multirow{2}{*}{Methods} 
& \multicolumn{3}{c}{EMGHandNet} 
& \multicolumn{3}{c}{Waveformer} \\
\cmidrule(lr){2-4} \cmidrule(lr){5-7}
& Acc & Macro-F1 & Macro-Rec 
& Acc & Macro-F1 & Macro-Rec \\
\midrule

Baseline 
& 70.28 $\pm$ 5.92 & 70.97 $\pm$ 5.49 & 71.26 $\pm$ 5.36
& 72.47 $\pm$ 6.20 & 73.11 $\pm$ 5.80 & 73.50 $\pm$ 5.70 \\

Replicate 
& 71.12 $\pm$ 5.90 & 71.91 $\pm$ 5.38 & 72.21 $\pm$ 5.26
& 73.35 $\pm$ 6.30 & 73.99 $\pm$ 5.90 & 74.33 $\pm$ 5.80 \\
\midrule
Jitter\&Scale 
& 70.63 $\pm$ 6.33 & 71.52 $\pm$ 5.86 & 71.73 $\pm$ 5.81
& 69.82 $\pm$ 6.90 & 70.63 $\pm$ 6.40 & 70.95 $\pm$ 6.30 \\

Upsample 
& 70.45 $\pm$ 6.14 & 71.18 $\pm$ 5.58 & 71.50 $\pm$ 5.48
& 73.59 $\pm$ 6.50 & 74.27 $\pm$ 6.00 & 74.68 $\pm$ 5.90 \\

Freq-Mask 
& 70.82 $\pm$ 5.98 & 71.56 $\pm$ 5.51 & 71.89 $\pm$ 5.41
& 71.63 $\pm$ 6.40 & 72.30 $\pm$ 5.90 & 72.70 $\pm$ 5.90 \\

Mixup 
& 73.09 $\pm$ 6.32 & 73.88 $\pm$ 5.79 & 74.14 $\pm$ 5.72
& 74.17 $\pm$ 6.80 & 75.01 $\pm$ 6.30 & 75.33 $\pm$ 6.20 \\

STAug
& 69.78 $\pm$ 6.29 & 70.49 $\pm$ 5.92 & 70.75 $\pm$ 5.79
& 72.82 $\pm$ 6.40 & 73.44 $\pm$ 6.00 & 73.80 $\pm$ 5.90 \\

Freq-Mix 
& 71.16 $\pm$ 5.85 & 71.96 $\pm$ 5.21 & 72.19 $\pm$ 5.20
& 72.15 $\pm$ 6.20 & 72.94 $\pm$ 5.70 & 73.32 $\pm$ 5.60 \\

\midrule

WGAN-GP~\cite{coelho2023novel} 
& 71.73 $\pm$ 6.10 & 72.59 $\pm$ 5.64 & 72.88 $\pm$ 5.54
& 73.97 $\pm$ 6.49 & 74.68 $\pm$ 6.01 & 75.02 $\pm$ 5.91 \\

DDPM~\cite{ho2020denoising}
& \textbf{74.23 $\pm$ 6.20} & \underline{74.92 $\pm$ 5.64} & \textbf{75.40 $\pm$ 5.54}
& \underline{75.53 $\pm$ 5.90} & 76.19 $\pm$ 5.49 & \textbf{76.73 $\pm$ 5.33} \\

PatchEMG~\cite{xiong2024patchemg}
& 70.52 $\pm$ 6.63 & 71.42 $\pm$ 6.03 & 71.72 $\pm$ 5.91
& 74.26 $\pm$ 6.50 & 75.02 $\pm$ 6.06 & 75.36 $\pm$ 5.90 \\

DDIM~\cite{song2020denoising} 
& 73.87 $\pm$ 5.56 & 74.67 $\pm$ 5.04 & 74.92 $\pm$ 4.92
& \textbf{75.61 $\pm$ 6.10} & \textbf{76.36 $\pm$ 5.60} & \underline{76.70 $\pm$ 5.50} \\

EMGFlow 
& \underline{74.19 $\pm$ 5.75} & \textbf{75.00 $\pm$ 5.21} & \underline{75.32 $\pm$ 5.14}
& 75.46 $\pm$ 6.10 & \underline{76.20 $\pm$ 5.70} & 76.60 $\pm$ 5.60 \\
\bottomrule
\end{tabular}
\end{table*}

\begin{table*}[t]
    \centering
    \caption{Train-on-synthetic test-on-real (TSTR) results of DDIM, DDPM, and EMGFlow across datasets and classifier backbones. In this protocol, the downstream classifier is trained only on synthetic data and then evaluated on held-out real test data, making it a stricter measure of standalone synthetic-data utility than standard augmentation. Results are reported as mean $\pm$ std (\%). EMGFlow consistently outperforms the accelerated DDIM baseline in all settings and is stronger than full 1000-step DDPM in five of the six dataset-backbone combinations.}
    \label{tab:tstr}
    \small
    \setlength{\tabcolsep}{5pt}
    \begin{tabular}{llcccc}
        \toprule
        Dataset & Backbone & Method & Accuracy $\uparrow$ & Macro-F1 $\uparrow$ & Macro-Recall $\uparrow$ \\
        \midrule
        \multirow{10}{*}{DB2} 
          & \multirow{5}{*}{EMGHandNet}
          & Baseline (real) & $70.28 \pm 5.92$ & $70.97 \pm 5.49$ & $71.26 \pm 5.36$ \\
          &  & WGAN-GP & $39.48 \pm 5.51$ & $40.25 \pm 5.54$ & $41.12 \pm 5.21$ \\
          &  & DDPM & $\mathbf{65.20 \pm 5.54}$ & $\mathbf{65.87 \pm 5.12}$ & $\mathbf{66.95 \pm 4.98}$ \\
          &  & DDIM & $62.16 \pm 6.43$ & $63.27 \pm 6.16$ & $63.92 \pm 5.94$ \\
          &  & EMGFlow   & $64.51 \pm 5.52$ & $65.50 \pm 5.16$ & $66.19 \pm 5.04$ \\
        \cmidrule(lr){2-6}
          & \multirow{5}{*}{Waveformer}
          & Baseline (real) & $72.50 \pm 6.20$ & $73.10 \pm 5.80$ & $73.50 \pm 5.70$ \\
          &  & WGAN-GP & $12.29 \pm 5.38$ & $10.80 \pm 4.68$ & $12.30 \pm 5.01$ \\
          &  & DDPM & $61.46 \pm 6.02$ & $61.89 \pm 5.78$ & $63.19 \pm 5.61$ \\
          &  & DDIM & $58.78 \pm 7.93$ & $59.87 \pm 7.55$ & $60.51 \pm 7.57$ \\
          &  & EMGFlow   & $\mathbf{61.93 \pm 5.77}$ & $\mathbf{62.69 \pm 5.36}$ & $\mathbf{63.44 \pm 5.27}$ \\
        \cmidrule(lr){1-6}
        \multirow{10}{*}{DB7} 
          & \multirow{5}{*}{EMGHandNet}
          & Baseline (real) & $73.05 \pm 4.48$ & $73.08 \pm 4.36$ & $73.38 \pm 4.37$ \\
          &  & WGAN-GP & $47.26 \pm 5.73$ & $46.97 \pm 5.84$ & $47.51 \pm 5.60$ \\
          &  & DDPM & $67.41 \pm 4.38$ & $67.62 \pm 4.22$ & $68.52 \pm 4.18$ \\
          &  & DDIM & $66.51 \pm 4.22$ & $67.03 \pm 4.02$ & $67.29 \pm 4.03$ \\
          &  & EMGFlow   & $\mathbf{68.91 \pm 4.89}$ & $\mathbf{69.33 \pm 4.59}$ & $\mathbf{69.95 \pm 4.40}$ \\
        \cmidrule(lr){2-6}
          & \multirow{5}{*}{Waveformer}
          & Baseline (real) & $76.36 \pm 3.81$ & $76.39 \pm 3.67$ & $76.62 \pm 3.63$ \\
          &  & WGAN-GP & $15.57 \pm 6.10$ & $13.65 \pm 5.52$ & $15.17 \pm 6.03$ \\
          &  & DDPM & $61.10 \pm 5.42$ & $61.00 \pm 5.41$ & $62.14 \pm 5.07$ \\
          &  & DDIM & $60.62 \pm 4.97$ & $60.85 \pm 4.86$ & $61.57 \pm 4.65$ \\
          &  & EMGFlow   & $\mathbf{64.16 \pm 4.43}$ & $\mathbf{64.15 \pm 4.29}$ & $\mathbf{65.06 \pm 4.15}$ \\
        \cmidrule(lr){1-6}
        \multirow{10}{*}{DB4} 
          & \multirow{5}{*}{EMGHandNet}
          & Baseline (real) & $64.74 \pm 5.38$ & $65.39 \pm 5.41$ & $65.65 \pm 5.60$ \\
          &  & WGAN-GP & $42.64 \pm 6.62$ & $42.82 \pm 7.01$ & $43.58 \pm 6.79$ \\
          &  & DDPM & $61.92 \pm 5.59$ & $62.61 \pm 5.84$ & $63.18 \pm 5.84$ \\
          &  & DDIM & $60.10 \pm 5.21$ & $61.38 \pm 5.25$ & $61.35 \pm 5.51$ \\
          &  & EMGFlow   & $\mathbf{62.83 \pm 5.87}$ & $\mathbf{63.69 \pm 5.62}$ & $\mathbf{64.10 \pm 5.98}$ \\
        \cmidrule(lr){2-6}
          & \multirow{5}{*}{Waveformer}
          & Baseline (real) & $68.72 \pm 5.76$ & $69.17 \pm 5.64$ & $69.54 \pm 5.75$ \\
          &  & WGAN-GP & $7.63 \pm 1.66$ & $6.57 \pm 1.52$ & $7.40 \pm 1.53$ \\
          &  & DDPM & $58.37 \pm 6.08$ & $58.94 \pm 6.19$ & $59.74 \pm 6.30$ \\
          &  & DDIM & $57.65 \pm 5.64$ & $58.57 \pm 5.61$ & $58.91 \pm 5.79$ \\
          &  & EMGFlow   & $\mathbf{60.55 \pm 6.21}$ & $\mathbf{61.22 \pm 6.14}$ & $\mathbf{61.80 \pm 6.40}$ \\
        \bottomrule
    \end{tabular}
\end{table*}

\section{Main Results}

\subsection{Comparison with Classical and Generative Augmentation Baselines}

We compare Flow Matching (FM) against a diverse set of augmentation baselines, including \textit{replicate}, jitter+scale~\cite{um2017data}, upsample~\cite{semenoglou2023data}, freq-mask~\cite{chen2023fraug}, mixup~\cite{zhang2018mixup}, STAug~\cite{zhang2023towards}, FreqMix~\cite{chen2023fraug}, WGAN-GP, PatchEMG~\cite{xiong2024patchemg}, DDIM, and DDPM. Here, DDIM denotes the DDPM-trained diffusion baseline evaluated with 50-step DDIM sampling, whereas DDPM denotes full 1000-step ancestral sampling. The \textit{replicate} baseline simply duplicates the training set once, serving as a control to separate the effect of increased sample count from the informational gain brought by synthetic data.

Table~\ref{tab:db7_aug}, Table~\ref{tab:db4_aug}, and Table~\ref{tab:db2_aug} reveal a consistent pattern. Learned generative augmentation is consistently stronger than most hand-crafted perturbation baselines, and full-step DDPM is generally better than accelerated DDIM, indicating that more expensive ancestral sampling can recover additional utility. WGAN-GP is often competitive with the stronger conventional baselines, but remains weaker than the best diffusion and EMGFlow results. PatchEMG, despite being designed for EMG augmentation, performs relatively weakly in our unified benchmark, suggesting that its patch-based design is not well aligned with the present setting.

Within this overall pattern, EMGFlow achieves the best or tied-best augmentation results in most dataset-backbone combinations and remains competitive even against full 1000-step DDPM. At the subject level, paired Wilcoxon signed-rank tests further show that EMGFlow significantly outperforms all conventional augmentation baselines as well as PatchEMG and WGAN-GP ($p<0.001$), whereas its differences relative to DDIM and full-step DDPM are not statistically significant. The main exception is DB2 with EMGHandNet, where DDPM is slightly better in ACC and Macro-Recall; however, this gain is small and comes with a much heavier sampling cost. Since augmentation is still anchored by real training data, the gap among strong generators is naturally narrower than in synthetic-only training. In this regime, EMGFlow's main advantage is therefore its stronger overall accuracy-efficiency trade-off rather than a large absolute margin on every metric.

\subsection{TSTR Results Across Generative Baselines.}

We further compare EMGFlow against both diffusion baselines, DDIM and full 1000-step DDPM, under TSTR, where the downstream classifier is trained on a class-balanced synthetic set with the same size as the real training set and evaluated on held-out real test data.

Table~\ref{tab:tstr} shows that EMGFlow provides the strongest overall standalone synthetic-data utility under the TSTR protocol. At the subject level, paired Wilcoxon signed-rank tests show that, on DB4 and DB7, EMGFlow significantly outperforms all competing generators ($p<0.001$); on DB2, it still significantly outperforms all methods except full-step DDPM ($p<0.001$). Accordingly, EMGFlow outperforms full 1000-step DDPM in five of the six dataset-backbone settings, with the only exception being DB2 with EMGHandNet. This exception should be interpreted together with the sampling budget: the DDPM result is obtained with roughly 25$\times$ the inference cost of EMGFlow, so the utility gain from long-run ancestral sampling is limited relative to its computational overhead. EMGFlow also consistently outperforms the accelerated DDIM baseline across all datasets and both backbones, and overall reaches roughly 84\%--97\% of the corresponding real-data baseline.

An additional observation is that TSTR performance depends more strongly on the downstream classifier than augmentation performance. Across all generators, Waveformer tends to be slightly weaker and more sensitive than EMGHandNet, suggesting that it is less tolerant to the residual distribution gap between synthetic and real data. A plausible explanation is that some generators match lower-order waveform statistics reasonably well while still deviating in higher-order temporal structure or class-discriminative feature geometry, to which Waveformer appears more sensitive. This effect is most pronounced for WGAN-GP: its TSTR results are already weaker on EMGHandNet and deteriorate further on Waveformer, which is consistent with the severe coverage deficiency and mode concentration typically associated with adversarial generation in this setting.

\subsection{Fidelity Metrics}

We next compare EMGFlow with both diffusion baselines, DDIM and DDPM, in terms of FID, IS, and CAS, using 256-dimensional features extracted by a pretrained EMGHandNet classifier and 5{,}000 samples for FID estimation. We additionally report the real train-versus-test baseline as a reference.

\begin{figure}[!t]
    \centering
    \includegraphics[width=\columnwidth]{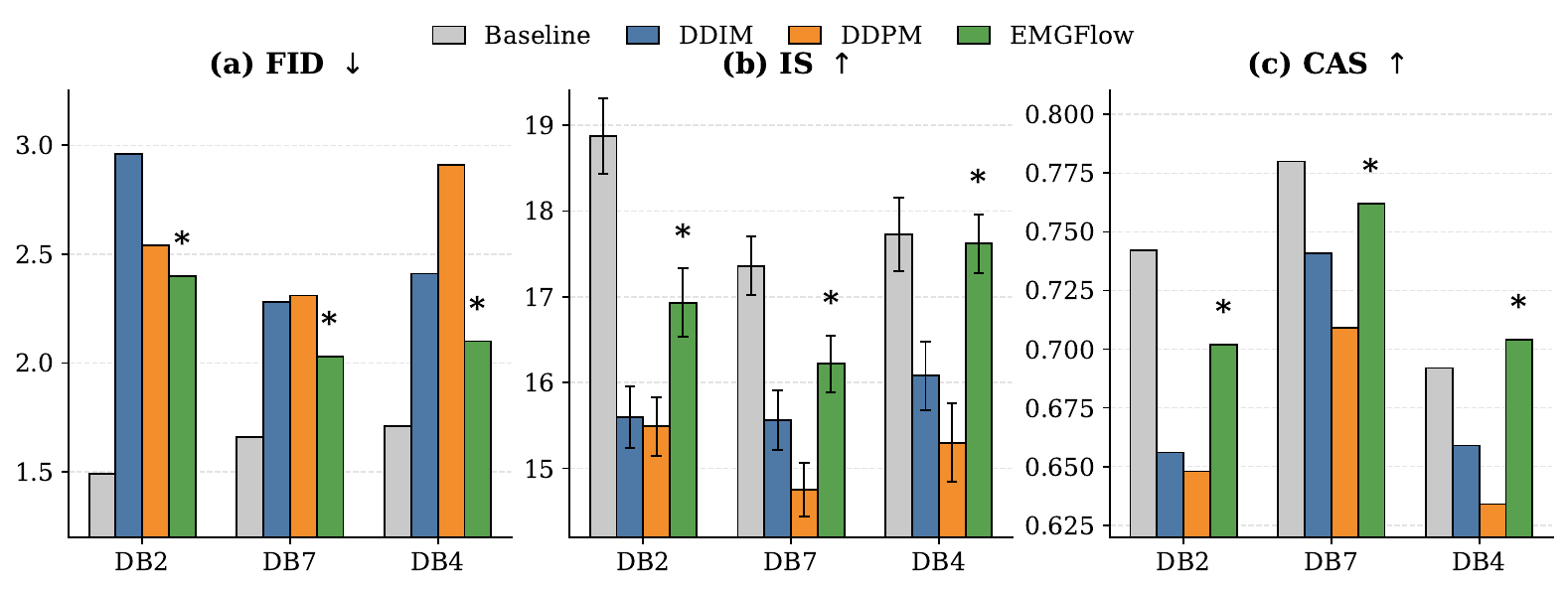}
    \caption{Single-column comparison of feature-based fidelity metrics across datasets. Each panel reports one metric, and the asterisk marks the best learned generator within each dataset. EMGFlow achieves the strongest overall fidelity, while the real-data baseline remains clearly better than all learned generators.}
    \label{fig:fidelity_bar_singlecol}
\end{figure}

\begin{figure*}[!t]
    \centering
    \begin{tabular}{ccc}
    \begin{subfigure}[]{
      \includegraphics[width=0.3\textwidth]{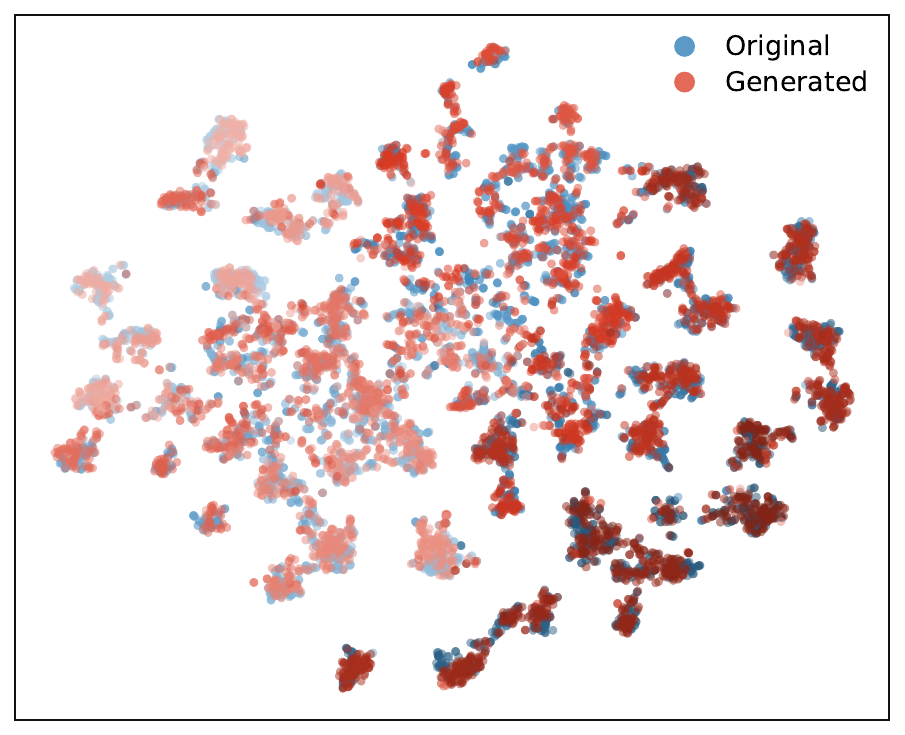}}
    \end{subfigure} &
    \begin{subfigure}[]{
      \includegraphics[width=0.3\textwidth]{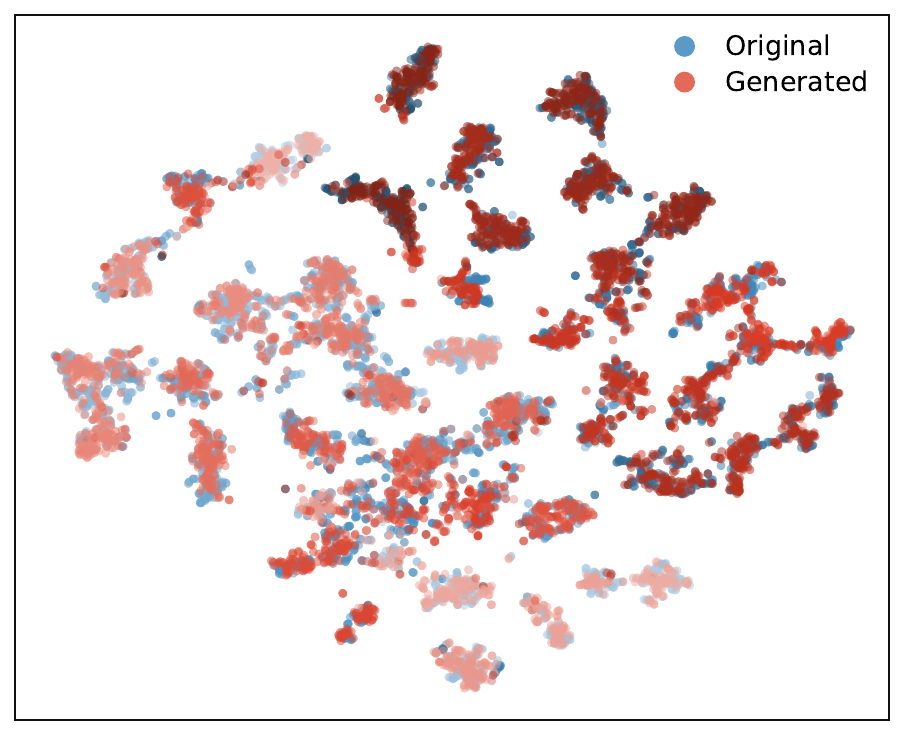}}
    \end{subfigure}
    \begin{subfigure}[]{
      \includegraphics[width=0.3\textwidth]{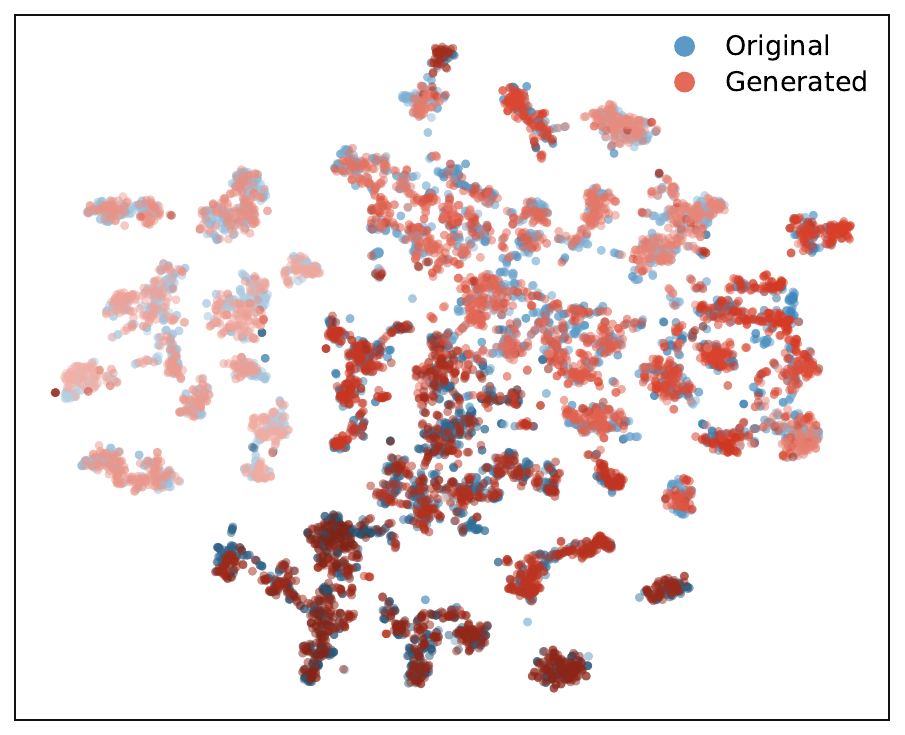}}
    \end{subfigure} &

  \end{tabular}
    \caption{t-SNE visualizations of real (blue) and generated (red) samples for one subject from (a) DB4, (b) DB7, and (c) DB2, where different shades denote different gesture classes. Across all three datasets, the generated points largely overlap the real class clusters, suggesting that EMGFlow captures the coarse class-conditional geometry in feature space. At the same time, several clusters still show mild shifts and density differences, which is consistent with the remaining realism gap indicated by the quantitative fidelity metrics.}
    \label{distribution}
\end{figure*}

\begin{figure*}[!t]
    \centering
    \subfigure[]{
        \includegraphics[width=0.48\textwidth]{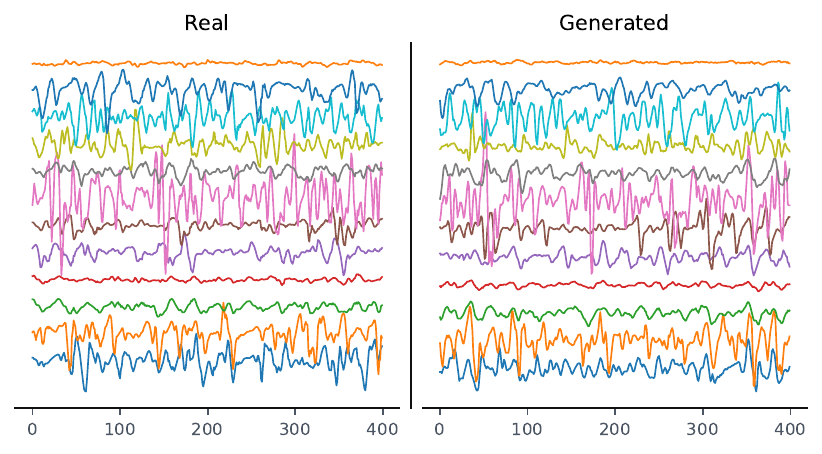}
    }
    \subfigure[]{
        \includegraphics[width=0.48\textwidth]{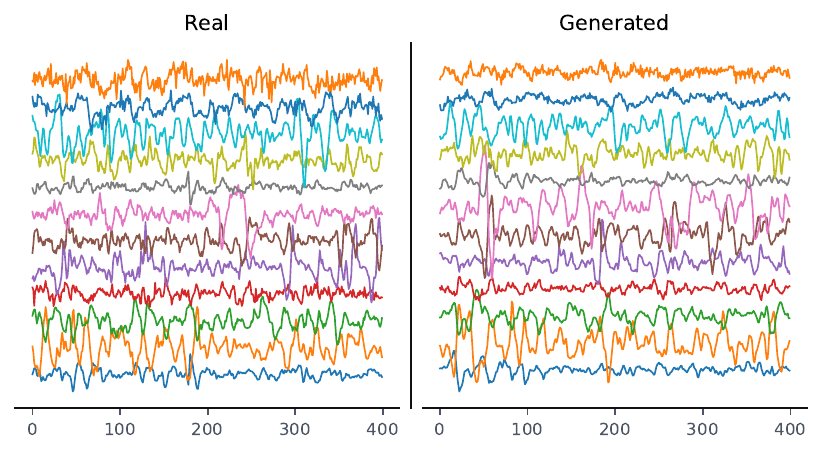}
    }
    \caption{Examples of real (left) and generated (right) multi-channel sEMG windows from (a) DB7 and (b) DB4. The generated signals reproduce the overall amplitude range, inter-channel variation, and transient burst structure of the corresponding real samples, while remaining visually distinct. These waveform-level examples qualitatively support the feature-space and downstream results, although slight differences in local smoothness and channel-specific details are still visible.}
    \label{samples}
\end{figure*}

Figure~\ref{fig:fidelity_bar_singlecol} shows that EMGFlow achieves the best overall feature-based fidelity among the learned generators across all three datasets. DDPM does not improve over DDIM on these metrics and is in fact weaker on DB4 and DB7, suggesting that the extra ancestral sampling cost does not directly translate into better feature alignment under this evaluation protocol. The real train-versus-test baseline remains better than all generative models, indicating that a substantial realism gap still persists. Figure~\ref{distribution} and Figure~\ref{samples} provide complementary feature-space and waveform-level examples.

\section{Empirical Analysis}

Beyond the main benchmark comparison, we further study several implementation choices that may materially affect conditional sEMG generation. These analyses cover both inference-time factors, such as guidance and solver selection, and training- or architecture-level factors, such as time sampling, conditioning interface, and normalization design, to clarify whether EMGFlow's gains rely on specific implementation details or reflect a more robust modeling advantage.

\subsection{Empirical Analysis of Classifier-Free Guidance}

Classifier-free guidance (CFG) is widely used in image generation to strengthen conditional consistency and often improve sample quality. However, its role in EMG generation remains unclear, especially when the goal is not only fidelity but also downstream utility for augmentation and synthetic training. To isolate the effect of guidance from other factors, we perform a dedicated scan on DB7 using the EMGHandNet backbone and vary the guidance weight over $w \in \{1.0, 1.25, 1.5, 2.0, 2.5\}$.

\begin{figure*}[!t]
    \centering
    \includegraphics[width=0.98\textwidth]{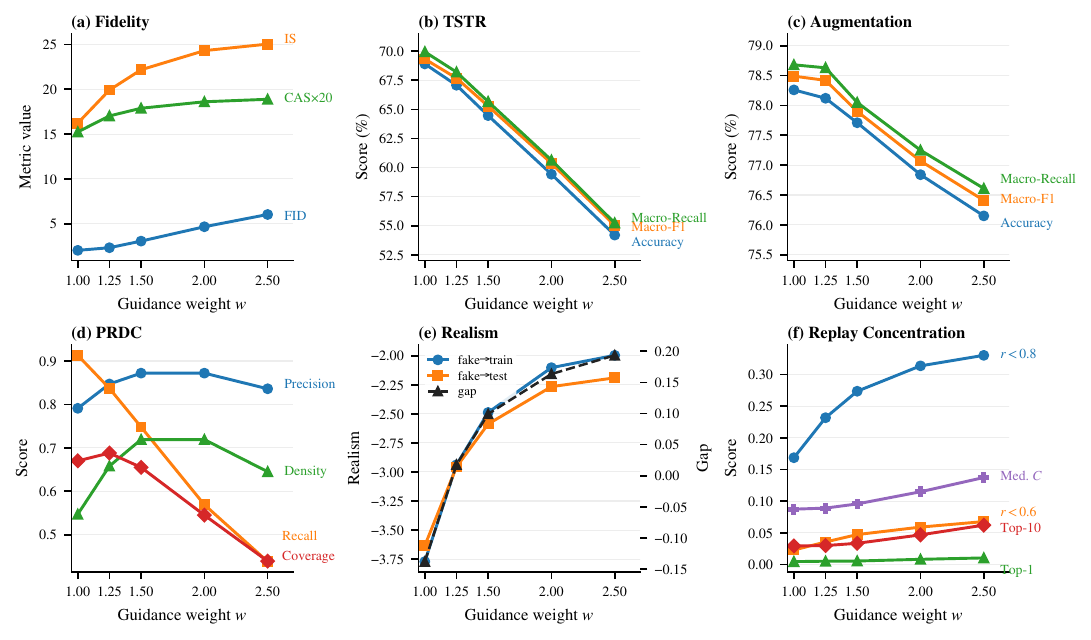}
    \caption{Summary of guidance effects on DB7 using EMGHandNet. (a) Feature-based fidelity metrics. (b) TSTR utility. (c) Augmentation utility. (d) PRDC metrics. (e) Neighborhood-based realism diagnostics, with the train-test gap shown on the right axis. (f) Prototype-concentration diagnostics. Stronger guidance improves class-discriminative and local-realism metrics, but reduces coverage and downstream utility.}
    \label{fig:guidance_summary}
\end{figure*}

Figure~\ref{fig:guidance_summary}(a) shows that increasing $w$ monotonically increases IS and CAS but worsens FID. Figure~\ref{fig:guidance_summary}(b) and Figure~\ref{fig:guidance_summary}(c) further show that stronger guidance consistently hurts downstream utility, with both TSTR and augmentation best at $w=1.0$.

To understand this mismatch between fidelity-like metrics and downstream performance, we next examine the local geometry of the generated distribution using PRDC. Figure~\ref{fig:guidance_summary}(d) shows that stronger guidance increases Precision and Density but reduces Recall and Coverage, indicating sharper prototypes but weaker support coverage. Thus, higher IS/CAS under strong guidance should not be interpreted as uniformly better generation quality.

To disentangle local realism from support coverage, we further introduce neighborhood-based diagnostics in the feature space: fake$\to$train realism, fake$\to$test realism, and the train-test gap. Figure~\ref{fig:guidance_summary}(e) shows that stronger guidance improves local realism but also increases train-set affinity.

Figure~\ref{fig:guidance_summary}(f) further shows a rising prototype concentration with stronger guidance. Taken together, these results suggest that stronger CFG sharpens prototypes and improves local realism, but reduces coverage and downstream utility; we therefore interpret this as a fidelity--utility tension with a \emph{prototype-like replay} tendency rather than strict copying.

This behavior is also broadly consistent with observations outside EMG: stronger guidance is often useful for perceptual sample quality, but lower guidance can be preferable when synthetic data are used for downstream model training~\cite{mei2024bigger,fan2024scaling}. In our setting, this suggests that increasing guidance contracts class-conditional support too aggressively, so gains in local realism are offset by losses in coverage and utility.

\subsection{Empirical Analysis of Solver Choice}
\label{sec:solver_analysis}

Flow Matching defines generation as solving a continuous-time ODE, so the choice of numerical solver can directly affect sample quality under a fixed inference budget. We therefore conduct a solver analysis on DB7 under matched numbers of function evaluations (NFE), first comparing FM with Heun against the DDIM diffusion baseline, and then comparing solvers within FM.

We first compare FM with Heun against the DDIM diffusion baseline under the same NFE budget. As shown in Table~\ref{tab:solver_ddim_heun}, Heun consistently outperforms DDIM on all metrics across the entire 10 to 50 NFE range; notably, FM with Heun at 10 NFE already achieves higher FID and CAS than DDIM at 50 NFE.

\begin{table}[!htbp]
\centering
\caption{Comparison between DDIM and Heun under matched numbers of function evaluations (NFE) on DB7. Under the same sampling budget, Heun consistently achieves lower FID, higher CAS, and higher IS than DDIM across the entire 10--50 NFE range. Notably, FM with Heun at 10 NFE already surpasses DDIM at 50 NFE on both FID and CAS, highlighting the efficiency advantage of Flow Matching with higher-order ODE solvers.} 
\label{tab:solver_ddim_heun}
\setlength{\tabcolsep}{6pt}
\begin{tabular}{c|cc|cc|cc}
\toprule
\multirow{2}{*}{NFE} 
& \multicolumn{2}{c|}{FID $\downarrow$} 
& \multicolumn{2}{c|}{CAS $\uparrow$} 
& \multicolumn{2}{c}{IS $\uparrow$} \\
& DDIM & Heun & DDIM & Heun & DDIM & Heun \\
\midrule
10 & 6.836 & \textbf{2.066} & 0.578 & \textbf{0.773} & 11.70 & \textbf{16.60} \\
20 & 3.359 & \textbf{2.006} & 0.686 & \textbf{0.764} & 14.27 & \textbf{16.26} \\
30 & 2.647 & \textbf{1.999} & 0.721 & \textbf{0.762} & 15.06 & \textbf{16.24} \\
40 & 2.407 & \textbf{1.998} & 0.733 & \textbf{0.758} & 15.42 & \textbf{16.14} \\
50 & 2.274 & \textbf{2.018} & 0.740 & \textbf{0.760} & 15.53 & \textbf{16.18} \\
\bottomrule
\end{tabular}

\end{table}

\begin{table}[!htbp]
\centering
\caption{Computational benchmark on Ninapro DB7. Since the diffusion baseline and EMGFlow use the same backbone, the comparison mainly highlights inference-side efficiency under different samplers and NFE budgets. EMGFlow with Heun achieves substantially lower FLOPs and higher throughput than DDPM and remains faster than DDIM, while retaining strong sample quality in the main experiments.}
\label{tab:benchmark_db7}
\begin{tabular}{lcccc}
\toprule
Method & NFE & FLOPs / Sample & Speed(sample/s)\\
\midrule
DDPM & 1000 & 4255.4M & 44.5 \\
DDIM & 50 & 208.57M & 902.4 \\
EMGFlow (Heun)   & 40 &  166.85M & 1117.4 \\
EMGFlow (Heun)   & 20 &  83.43M  & 2225.6 \\
\bottomrule
\end{tabular}
\end{table}

This advantage is also conceptually consistent with the modeling difference between the two samplers. DDIM is an accelerated sampler derived from a diffusion process trained on a discrete noising schedule, whereas FM directly learns a continuous velocity field and generates by integrating the corresponding ODE. As a result, higher-order numerical solvers can exploit the local structure of the learned FM vector field more directly, which helps explain why FM benefits more visibly from improved solver accuracy under matched NFE.

\begin{figure}[!htbp]
    \centering
    \subfigure[DB4]{
        \includegraphics[height=0.33\textwidth,trim=0 0 212 20,clip]{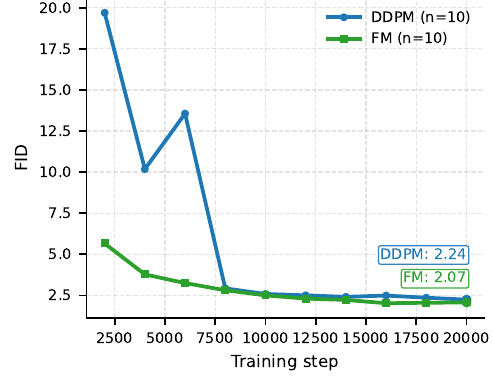}
    }
    \hspace{0.06\textwidth}
    \subfigure[DB7]{
        \includegraphics[height=0.33\textwidth,trim=235 0 0 20,clip]{fig6.pdf}
    }
    \vspace{-0.02\textwidth}

    \caption{FID trajectories of the diffusion baseline and EMGFlow during training on DB4 and DB7. EMGFlow reduces FID more rapidly in the early stage and reaches a stable low-FID regime with fewer training steps on both datasets. This trend suggests that Flow Matching is not only efficient at inference time, but also easier to optimize under the present training setup.}
    \label{fig:fid_history_ddpm_vs_fm_db4_db7}
\end{figure}

To complement the matched-NFE comparison, we also examine practical efficiency. Since the diffusion baseline and FM share the same backbone, their training cost is also approximately matched. Table~\ref{tab:benchmark_db7} therefore mainly highlights the inference-side advantage of FM, while Figure~\ref{fig:fid_history_ddpm_vs_fm_db4_db7} shows that FM also reaches low FID earlier on both DB4 and DB7.

\begin{figure*}[!htbp]
    \centering
    \includegraphics[width=0.95\textwidth]{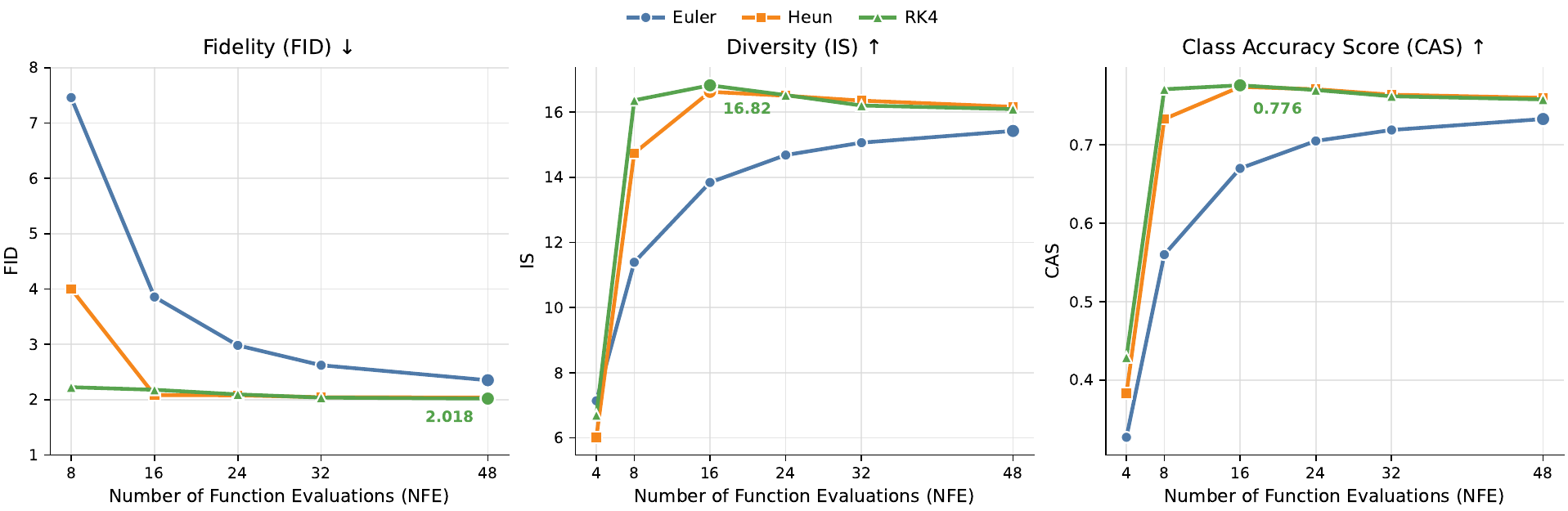}
    \caption{Comparison of Euler, Heun, and RK4 within EMGFlow under matched numbers of function evaluations (NFE) on DB7. When the sampling budget becomes moderately large, Heun and RK4 consistently achieve lower FID and higher CAS than Euler, showing that higher-order integration better exploits the learned continuous vector field. Euler is only comparatively less poor in the extremely low-NFE regime, where the number of effective integration steps is too small for multi-stage solvers to realize their accuracy advantage.}
    \label{fig:solver_fm_solvers}
\end{figure*}

We next compare three ODE solvers within FM itself, namely Euler, Heun, and RK4, again under matched NFE. Figure~\ref{fig:solver_fm_solvers} reveals a clear regime-dependent pattern.

At extremely small budgets (e.g., 4 NFE), Euler only appears relatively less degraded than Heun and RK4. This behavior is expected under a fixed-NFE comparison, because higher-order solvers consume multiple function evaluations per step. Under such a coarse discretization, the higher-order methods do not yet have enough steps to realize their accuracy advantage, and all three solvers remain in a poor regime overall. At this point, the dominant error source is not only the formal local truncation order, but also the fact that the trajectory is simply sampled too sparsely for multi-stage corrections to unfold effectively. At 4 NFE, the corresponding FID values are 19.637 for Euler, 33.006 for Heun, and 30.370 for RK4. The latter two values are omitted from Figure~\ref{fig:solver_fm_solvers} to preserve the readability of the main FID range.

Once the NFE budget becomes moderately large ($\ge 8$ NFE), however, higher-order solvers rapidly overtake Euler, delivering substantially lower FID and higher CAS. From 16 NFE onward, Heun and RK4 remain consistently stronger than Euler, while their mutual difference becomes relatively small.

\subsection{Empirical Analysis of Time Sampling}
\label{sec:tsampling_analysis}
In our main experiments, Flow Matching uses logit-normal time sampling as the default configuration. To assess this design choice, we replace the time sampling distribution $p(t)$ in Eq.~(\ref{eq:fm_objective}) with uniform sampling over $[0,1]$.

The intuition behind logit-normal sampling is that not all parts of the trajectory are equally informative during training. Very small $t$ corresponds to states that remain close to noise and therefore carry weak class-specific structure, while very large $t$ is already close to the data manifold and may provide comparatively redundant supervision. The middle portion of the trajectory more directly constrains how class-dependent structure emerges from noise, so emphasizing this regime can improve the learned transport behavior.

\begin{figure*}[!htbp]
\centering
\includegraphics[width=0.98\textwidth]{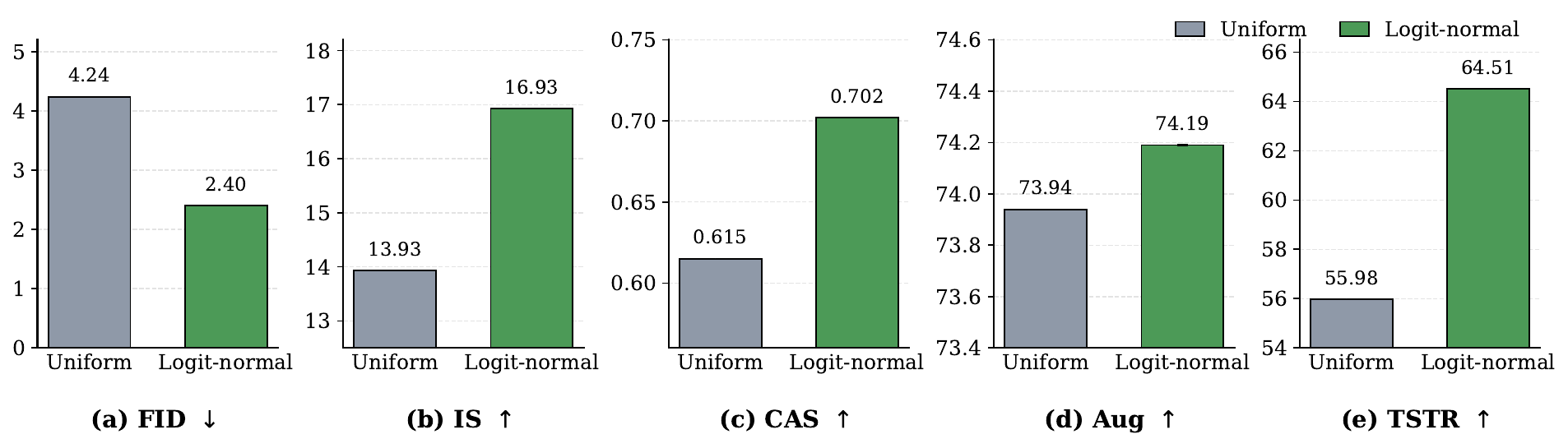}
\caption{Effect of time sampling strategy on DB2. The five panels report FID, IS, CAS, augmentation accuracy, and TSTR accuracy, respectively. Compared with uniform sampling, logit-normal time sampling improves all reported metrics, with the largest gain appearing under the more stringent TSTR setting. This result suggests that emphasizing the intermediate portion of the flow trajectory provides a more useful training bias than sampling time uniformly.}
\label{fig:tsampling_db2}
\end{figure*}

Figure~\ref{fig:tsampling_db2} shows that uniform sampling degrades both fidelity and downstream utility on DB2, especially under TSTR. This suggests that, in our setting, allocating more supervision to the intermediate regime is beneficial not only for feature-space alignment but also for the standalone utility of the generated signals.

At the same time, logit-normal sampling is more optimization-sensitive in practice, with cosine annealing helping stabilize training. A plausible explanation is that non-uniform $t$ sampling changes the gradient budget assigned to different difficulty regimes along the trajectory, making optimization more dependent on learning-rate scheduling and EMA smoothing.

It is also worth clarifying the objective-level implication of this design. If one starts from the standard uniform-time FM objective and samples $t$ from an alternative proposal distribution $q(t)$, then importance weighting by $p_{\mathrm{uni}}(t)/q(t)$ would be required to preserve the original objective. This perspective has also been emphasized more broadly in diffusion design analyses~\cite{karras2022elucidating}. In our experiments, however, we do not apply such importance weighting. The logit-normal variant should therefore be interpreted as deliberately reweighting the training objective toward the middle of the trajectory, rather than as an unbiased estimator of the uniform objective. Interestingly, in small-scale pilot experiments, adding importance weights made the results nearly revert to the uniform baseline, whereas the unweighted logit-normal objective performed best. Although this observation deserves further theoretical study, it suggests that the gain may come not only from sampling efficiency but also from the inductive bias introduced by changing which parts of the transport path receive more emphasis.

\subsection{Empirical Analysis of Conditioning and Normalization Design}
\label{sec:condnorm_analysis}

In our default architecture, class information is injected through adaptive GroupNorm-style modulation. To examine whether EMGFlow's gains depend on this design, we further ablate the conditioning interface and normalization strategy on DB7 using the EMGHandNet backbone. Preliminary runs that revert GroupNorm to the original BatchNorm-based design are highly unstable: the FID trajectory oscillates sharply during training, and several subjects fail to converge reliably. We therefore exclude BatchNorm from the formal quantitative comparison and focus on three stable GroupNorm-based variants: the default GN+AdaGN, GN+add, and GN+concat.

Table~\ref{tab:condnorm_downstream_db7} shows that the default GN+AdaGN remains best under both augmentation and TSTR. Relative to the default design, GN+add leads to statistically significant drops in all downstream metrics under both settings (all $p<0.005$), while GN+concat degrades even more severely (all $p<10^{-4}$). These conclusions are based on paired Wilcoxon signed-rank tests over 20 subjects. This indicates that keeping the same normalized backbone is not sufficient; the way class information is injected has a direct impact on the usefulness of the generated data.

\begin{table}[!htbp]
\centering
\caption{Effect of conditioning interface on downstream utility on DB7 using EMGHandNet. Results are reported as mean $\pm$ std (\%). The default GN+AdaGN design achieves the best performance under both augmentation and TSTR, while simpler GN+add and GN+concat variants lead to consistent degradation, showing that the conditioning interface materially affects the usefulness of the generated data rather than serving as a minor implementation detail.}
\label{tab:condnorm_downstream_db7}
\small
\setlength{\tabcolsep}{5pt}
\begin{tabular}{lcccccc}
\toprule
\multirow{2}{*}{Method}
& \multicolumn{3}{c}{Augmentation}
& \multicolumn{3}{c}{TSTR} \\
\cmidrule(lr){2-4} \cmidrule(lr){5-7}
& Acc & Macro-F1 & Macro-Rec
& Acc & Macro-F1 & Macro-Rec \\
\midrule
GN+AdaGN (default)
& \textbf{78.26 $\pm$ 3.95} & \textbf{78.49 $\pm$ 3.69} & \textbf{78.68 $\pm$ 3.71}
& \textbf{68.91 $\pm$ 4.89} & \textbf{69.33 $\pm$ 4.59} & \textbf{69.95 $\pm$ 4.40} \\
GN+add
& 77.36 $\pm$ 3.50 & 77.54 $\pm$ 3.39 & 77.69 $\pm$ 3.31
& 67.45 $\pm$ 4.32 & 67.80 $\pm$ 4.09 & 68.22 $\pm$ 3.98 \\
GN+concat
& 77.15 $\pm$ 3.93 & 77.33 $\pm$ 3.88 & 77.60 $\pm$ 3.80
& 64.62 $\pm$ 5.11 & 64.93 $\pm$ 4.63 & 65.47 $\pm$ 4.59 \\
\bottomrule
\end{tabular}
\end{table}

From the fidelity perspective, GN+add remains relatively close to the default model in FID (2.03 vs.\ 2.12; $p=0.07$), but both IS and CAS already decrease significantly (both $p<0.005$), suggesting that simple additive conditioning can partially preserve global distribution alignment while still weakening class-aware structure modeling. GN+concat is significantly worse on FID, IS, and CAS (all $p<10^{-4}$), indicating that naive concatenation is a poor conditional interface in this setting.

\begin{table}[!htbp]
\centering
\caption{Effect of conditioning interface on fidelity and distributional geometry on DB7 using EMGHandNet. The default GN+AdaGN design achieves the best overall balance of FID, IS, CAS, and PRDC-style geometry metrics, while simpler GN+add and GN+concat variants degrade class-discriminative fidelity and coverage. This result suggests that adaptive conditional normalization improves class-aware generation without collapsing support.}
\label{tab:condnorm_fidelity_db7}
\small
\renewcommand{\arraystretch}{1.15}
\setlength{\tabcolsep}{7pt}
\begin{tabular}{lccccccc}
\toprule
Method & FID $\downarrow$ & IS $\uparrow$ & CAS $\uparrow$ & PR $\uparrow$ & RE $\uparrow$ & Density $\uparrow$ & Coverage $\uparrow$ \\
\midrule
GN+AdaGN (default) & \textbf{2.030} & \textbf{16.217 $\pm$ 0.327} & \textbf{0.762} & \textbf{0.791} & 0.913 & \textbf{0.547} & \textbf{0.670} \\
GN+add             & 2.115 & 15.518 $\pm$ 0.316 & 0.736 & 0.762 & 0.931 & 0.505 & 0.646 \\
GN+concat          & 2.697 & 13.480 $\pm$ 0.286 & 0.660 & 0.723 & \textbf{0.940} & 0.446 & 0.589 \\
\bottomrule
\end{tabular}
\end{table}

The PRDC statistics in Table~\ref{tab:condnorm_fidelity_db7} further clarify the mechanism. Compared with both GN+add and GN+concat, the default GN+AdaGN achieves significantly higher precision, density, and coverage (all $p<0.001$). Although the default model has slightly lower recall than both simpler variants, the absolute difference is small, and its higher coverage indicates that the better IS and CAS are not obtained by collapsing support.

This behavior is notably different from the guidance effect analyzed earlier. Stronger CFG improved IS and CAS mainly by contracting class-conditional support and increasing prototype concentration, whereas the default GN+AdaGN design improves downstream utility and class-discriminative fidelity while retaining better coverage than simpler conditioning interfaces. We therefore interpret the benefit of adaptive conditional normalization not as stronger prototype replay, but as more effective class-conditional structure modeling without sacrificing intra-class support.

\section{Discussion}

\subsection{TSTR as a Stricter Indicator of Standalone Utility}

The contrast between augmentation and TSTR is itself informative about what a generator is actually contributing. In augmentation, synthetic samples are added on top of real training data, so their role is mainly to densify the observed distribution, fill local decision-boundary regions, and increase the effective sample count. Under this setting, the real data still anchor the class structure, and even a moderately effective generator can be helpful. TSTR is more demanding: the synthetic set must define the entire training distribution on its own, without any real-data support. It therefore tests whether the generator has learned a class-conditional distribution that transfers to real data, rather than simply whether its samples are locally useful when real examples already stabilize the classifier.

This distinction helps explain why EMGFlow is only competitive with strong diffusion baselines under augmentation, yet more clearly advantageous under TSTR. Augmentation accuracy answers whether synthetic data can help; TSTR more directly answers whether the generator has learned a transferable distribution. For this reason, we view TSTR as a stricter indicator of standalone utility, and we argue that augmentation performance alone is insufficient for judging the true quality of a synthetic-data generator.

\subsection{Earlier Fidelity--Utility Tension Under Stronger Guidance in sEMG}

Our guidance analysis suggests a sharper tension between local realism and useful support coverage than is commonly emphasized in natural-image generation. In image domains, stronger classifier-free guidance can often improve perceptual quality while leaving sufficient intra-class variation intact for downstream use. We hypothesize that this earlier fidelity--utility tension in sEMG arises because the class-conditional support is intrinsically narrower, while the overlapping sliding-window protocol further increases local redundancy among neighboring samples. Under these conditions, stronger guidance may sharpen high-density class prototypes faster than it preserves the full intra-class support.

This interpretation is consistent with the joint behavior of our metrics: stronger guidance raises IS, CAS, precision, density, local realism, and prototype concentration, yet lowers recall, coverage, TSTR utility, and augmentation utility. In other words, local samples become more class-typical, but the generated distribution becomes less useful as a training set. We therefore interpret the guidance results as being consistent with the view that, in sEMG generation, stronger conditional sharpening may improve prototype realism faster than it preserves useful class-conditional coverage, leading to an earlier fidelity--utility tension than one might expect from natural-image generation.

\subsection{Interpreting Feature-Based Fidelity for Physiological Signal Generation}

Our results also clarify how feature-based fidelity metrics should be interpreted in physiological signal generation. FID measures distribution alignment in a learned feature space and is therefore useful for detecting broad realism gaps and severe support collapse; unlike IS or CAS, it is not improved merely by making samples more class-confident. However, FID is still not a task-oriented endpoint: its value depends on the chosen feature extractor, and good feature-space alignment does not automatically imply maximal downstream usefulness.

By contrast, IS and CAS are more sensitive to sample-level discriminability and classifier confidence. These properties make them useful diagnostics, but also potentially misleading when support becomes overly concentrated, as seen in the guidance analysis. For task-oriented biosignals such as sEMG, the ultimate goal is usually not perceptual quality in itself, but whether synthetic data improve downstream recognition, robustness, or generalization. We therefore regard feature-based fidelity as an informative proxy rather than a sufficient criterion, and argue that synthetic physiological signals should be evaluated jointly in terms of fidelity, coverage, downstream utility, and efficiency.

\subsection{Limitations and Future Work}

While this study demonstrates the potential of Flow Matching (FM) for sEMG generation and augmentation, several limitations remain. First, our experiments follow a within-subject, cross-trial protocol; whether generative augmentation improves cross-subject or cross-session generalization remains an open question. Second, although we evaluate on three public Ninapro benchmarks, broader validation across more diverse datasets and acquisition conditions is still needed. Third, generation is performed at the fixed-window level to match the downstream classification pipeline, rather than at the full-trial level. Finally, all feature-based fidelity metrics are computed using a fixed pretrained EMGHandNet feature extractor, so their absolute values should be interpreted within this protocol.

Future work will extend FM to more challenging cross-subject and cross-session settings, explore faster low-NFE or distilled sampling strategies, and investigate whether similar observations hold for other physiological time series such as EEG and ECG.
\section{Conclusion}
In this work, we presented EMGFlow, a Flow Matching based framework for conditional sEMG generation and data augmentation. Across three Ninapro benchmarks, EMGFlow consistently outperformed conventional augmentation methods and remained stronger than or comparable to representative generative baselines, including WGAN-GP and DDIM. Its advantage was particularly evident under the stricter TSTR setting, where FM provided stronger standalone synthetic-data utility.

Beyond downstream recognition, FM also achieved better or comparable feature-based fidelity than DDIM, while solver analysis showed an efficiency advantage under matched inference budgets. Our empirical studies further showed that practical FM design choices matter: stronger classifier-free guidance improves class-discriminative metrics but can reduce coverage and downstream usefulness, adaptive conditional normalization is more effective than simpler add or concat interfaces, logit-normal time sampling is more effective than uniform sampling, and Heun provides a favorable trade-off between quality and efficiency in our setting.

Taken together, these results support Flow Matching as a strong and practical generative baseline for EMG augmentation, while also showing that synthetic-data evaluation should jointly consider fidelity, coverage, utility, and efficiency rather than relying on any single metric alone.

\section*{Declaration of competing interest}
The authors declare that they have no known competing financial interests or personal relationships that could have appeared to influence the work reported in this paper.

\section*{Declaration of generative AI and AI-assisted technologies in the writing process}
During the preparation of this work the authors used ChatGPT in order to improve language and readability. After using this tool/service, the authors reviewed and edited the content as needed and take full responsibility for the content of the publication.

\bibliographystyle{elsarticle-num}
\bibliography{refs}

@article{mei2024bigger,
  title={Bigger is not Always Better: Scaling Properties of Latent Diffusion Models},
  author={Mei, Kangfu and Tu, Zhengzhong and Delbracio, Mauricio and Talebi, Hossein and Patel, Vishal M. and Milanfar, Peyman},
  journal={Transactions on Machine Learning Research},
  year={2025},
  note={Accepted in 2024; arXiv:2404.01367}
}

@inproceedings{fan2024scaling,
  title={Scaling Laws of Synthetic Images for Model Training ... for Now},
  author={Fan, Lijie and Chen, Kaifeng and Krishnan, Dilip and Katabi, Dina and Isola, Phillip and Tian, Yonglong},
  booktitle={Proceedings of the IEEE/CVF Conference on Computer Vision and Pattern Recognition},
  pages={7382--7392},
  year={2024}
}

@article{kaifosh2025generic,
  title={A generic non-invasive neuromotor interface for human-computer interaction},
  author={Kaifosh, Patrick and Reardon, Thomas R},
  journal={Nature},
  pages={1--10},
  year={2025},
  publisher={Nature Publishing Group UK London}
}

@inproceedings{dhariwal2021diffusion,
  title={Diffusion models beat gans on image synthesis},
  author={Dhariwal, Prafulla and Nichol, Alexander},
  booktitle={Advances in Neural Information Processing Systems},
  volume={34},
  pages={8780--8794},
  year={2021}
}

@inproceedings{karras2022elucidating,
  title = {Elucidating the Design Space of Diffusion-Based Generative Models},
  booktitle = {Advances in Neural Information Processing Systems},
  author = {Karras, Tero and Aittala, Miika and Aila, Timo and Laine, Samuli},
  year = 2022
}

@article{cao2024survey,
  title={A survey of mix-based data augmentation: Taxonomy, methods, applications, and explainability},
  author={Cao, Chengtai and Zhou, Fan and Dai, Yurou and Wang, Jianping and Zhang, Kunpeng},
  journal={ACM Computing Surveys},
  volume={57},
  number={2},
  pages={1--38},
  year={2024},
  publisher={ACM New York, NY}
}

@article{zhao2024dominant,
  title={Dominant Shuffle: A Simple Yet Powerful Data Augmentation for Time-series Prediction},
  author={Zhao, Kai and He, Zuojie and Hung, Alex and Zeng, Dan},
  journal={arXiv preprint arXiv:2405.16456},
  year={2024}
}

@article{xiong2024patchemg,
  title={{PatchEMG}: Few-shot {EMG} signal generation with diffusion models for data augmentation to improve classification performance},
  author={Xiong, Baoping and Chen, Wensheng and Li, Han and Niu, Yinxi and Zeng, Nianyin and Gan, Zhenhua and Xu, Yong},
  journal={IEEE Transactions on Instrumentation and Measurement},
  year={2024},
  publisher={IEEE}
}

@article{esteban2017real,
  title   = {Real-Valued (Medical) Time Series Generation with Recurrent Conditional {GAN}s},
  author  = {Esteban, Crist{\'o}bal and Hyland, Stephanie L. and R{\"a}tsch, Gunnar},
  journal = {arXiv preprint arXiv:1706.02633},
  year    = {2017},
  eprint  = {1706.02633},
  archivePrefix = {arXiv},
  primaryClass  = {cs.LG}
}

@article{atzori2014electromyography,
  title={Electromyography data for non-invasive naturally-controlled robotic hand prostheses},
  author={Atzori, Manfredo and Gijsberts, Arjan and Castellini, Claudio and Caputo, Barbara and Hager, Anne-Gabrielle Mittaz and Elsig, Simone and Giatsidis, Giorgio and Bassetto, Franco and M{\"u}ller, Henning},
  journal={Scientific Data},
  volume={1},
  number={1},
  pages={1--13},
  year={2014},
  publisher={Nature Publishing Group}
}

@article{pizzolato2017comparison,
  title={Comparison of six electromyography acquisition setups on hand movement classification tasks},
  author={Pizzolato, Stefano and Tagliapietra, Luca and Cognolato, Matteo and Reggiani, Monica and M{\"u}ller, Henning and Atzori, Manfredo},
  journal={PLoS ONE},
  volume={12},
  number={10},
  pages={e0186132},
  year={2017},
  publisher={Public Library of Science San Francisco, CA USA}
}

@article{krasoulis2017improved,
  title={Improved prosthetic hand control with concurrent use of myoelectric and inertial measurements},
  author={Krasoulis, Agamemnon and Kyranou, Iris and Erden, Mustapha Suphi and Nazarpour, Kianoush and Vijayakumar, Sethu},
  journal={Journal of NeuroEngineering and Rehabilitation},
  volume={14},
  number={1},
  pages={71},
  year={2017},
  publisher={Springer}
}

@article{wang2024transformer,
  title={Transformer-based network with temporal depthwise convolutions for {sEMG} recognition},
  author={Wang, Zefeng and Yao, Junfeng and Xu, Meiyan and Jiang, Min and Su, Jinsong},
  journal={Pattern Recognition},
  volume={145},
  pages={109967},
  year={2024},
  publisher={Elsevier}
}

@article{al2024tcnn,
  title={TCNN-KAN: Optimized CNN by Kolmogorov-Arnold network and pruning techniques for {sEMG} gesture recognition},
  author={Al-Qaness, Mohammed AA and Ni, Sike},
  journal={IEEE Journal of Biomedical and Health Informatics},
  volume={29},
  number={1},
  pages={188--197},
  year={2025},
  doi={10.1109/JBHI.2024.3467065},
  publisher={IEEE}
}

@inproceedings{heusel2017gans,
  title={Gans trained by a two time-scale update rule converge to a local nash equilibrium},
  author={Heusel, Martin and Ramsauer, Hubert and Unterthiner, Thomas and Nessler, Bernhard and Hochreiter, Sepp},
  booktitle={Advances in Neural Information Processing Systems},
  volume={30},
  year={2017}
}

@article{yang2025stcnet,
  title={STCNet: Spatio-Temporal Cross Network with subject-aware contrastive learning for hand gesture recognition in surface {EMG}},
  author={Yang, Jaemo and Cha, Doheun and Lee, Dong-Gyu and Ahn, Sangtae},
  journal={Computers in Biology and Medicine},
  volume={185},
  pages={109525},
  year={2025},
  publisher={Elsevier}
}

@inproceedings{um2017data,
  title={Data augmentation of wearable sensor data for parkinson’s disease monitoring using convolutional neural networks},
  author={Um, Terry T and Pfister, Franz MJ and Pichler, Daniel and Endo, Satoshi and Lang, Muriel and Hirche, Sandra and Fietzek, Urban and Kuli{\'c}, Dana},
  booktitle={Proceedings of the ACM International Conference on Multimodal Interaction},
  pages={216--220},
  year={2017}
}

@article{semenoglou2023data,
  title={Data augmentation for univariate time series forecasting with neural networks},
  author={Semenoglou, Artemios-Anargyros and Spiliotis, Evangelos and Assimakopoulos, Vassilios},
  journal={Pattern Recognition},
  volume={134},
  pages={109132},
  year={2023},
  publisher={Elsevier}
}

@article{chen2023fraug,
  title={FrAug: Frequency domain augmentation for time series forecasting},
  author={Chen, Muxi and Xu, Zhijian and Zeng, Ailing and Xu, Qiang},
  journal={arXiv preprint arXiv:2302.09292},
  year={2023}
}

@inproceedings{zhang2018mixup,
  title={mixup: Beyond Empirical Risk Minimization},
  author={Zhang, Hongyi and Cisse, Moustapha and Dauphin, Yann N and Lopez-Paz, David},
  booktitle={International Conference on Learning Representations},
  year={2018}
}

@inproceedings{zhang2023towards,
  title={Towards diverse and coherent augmentation for time-series forecasting},
  author={Zhang, Xiyuan and Chowdhury, Ranak Roy and Shang, Jingbo and Gupta, Rajesh and Hong, Dezhi},
  booktitle={Proceedings of the IEEE International Conference on Acoustics, Speech and Signal Processing},
  pages={1--5},
  year={2023},
  organization={IEEE}
}

@inproceedings{ho2021classifier,
  title={Classifier-Free Diffusion Guidance},
  author={Ho, Jonathan and Salimans, Tim},
  booktitle={Advances in Neural Information Processing Systems Workshop},
  year = {2021}
}

@article{tsinganos2020data,
  title={Data augmentation of surface electromyography for hand gesture recognition},
  author={Tsinganos, Panagiotis and Cornelis, Bruno and Cornelis, Jan and Jansen, Bart and Skodras, Athanassios},
  journal={Sensors},
  volume={20},
  number={17},
  pages={4892},
  year={2020},
  publisher={MDPI}
}

@article{jiang2022optimization,
  title = {Optimization of HD-sEMG-Based Cross-Day Hand Gesture Classification by Optimal Feature Extraction and Data Augmentation},
  author = {Jiang, Xinyu and Liu, Xiangyu and Fan, Jiahao and Ye, Xinming and Dai, Chenyun and Clancy, Edward A. and Farina, Dario and Chen, Wei},
  year = {2022},
  journal = {IEEE Transactions on Human-Machine Systems},
  volume = {52},
  number = {6},
  pages = {1281--1291},
  issn = {2168-2291, 2168-2305},
  doi = {10.1109/THMS.2022.3175408},
  copyright = {https://ieeexplore.ieee.org/Xplorehelp/downloads/license-information/IEEE.html}
}

@article{meng2022usertailored,
  title = {User-Tailored Hand Gesture Recognition System for Wearable Prosthesis and Armband Based on Surface Electromyogram},
  author = {Meng, Long and Jiang, Xinyu and Liu, Xiangyu and Fan, Jiahao and Ren, Haoran and Guo, Yao and Diao, Haikang and Wang, Zihao and Chen, Chen and Dai, Chenyun and Chen, Wei},
  year = {2022},
  journal = {IEEE Transactions on Instrumentation and Measurement},
  volume = {71},
  pages = {2520616},
  doi = {10.1109/TIM.2022.3217868}
}

@article{bird2021synthetic,
  title={Synthetic biological signals machine-generated by GPT-2 improve the classification of EEG and EMG through data augmentation},
  author={Bird, Jordan J and Pritchard, Michael and Fratini, Antonio and Ek{\'a}rt, Anik{\'o} and Faria, Diego R},
  journal={IEEE Robotics and Automation Letters},
  volume={6},
  number={2},
  pages={3498--3504},
  year={2021},
  publisher={IEEE}
}

@inproceedings{mendez2022emg,
  title={EMG data augmentation for grasp classification using generative adversarial networks},
  author={Mendez, Vincent and Lhoste, Cl{\'e}ment and Micera, Silvestro},
  booktitle={Proceedings of the Annual International Conference of the IEEE Engineering in Medicine and Biology Society},
  pages={3619--3622},
  year={2022},
  organization={IEEE}
}

@inproceedings{lipman2022flow,
  title = {Flow Matching for Generative Modeling},
  booktitle = {International Conference on Learning Representations},
  author = {Lipman, Yaron and Chen, Ricky T. Q. and {Ben-Hamu}, Heli and Nickel, Maximilian and Le, Matthew},
  year = 2022
}

@inproceedings{liu2022flow,
  title = {Flow Straight and Fast: Learning to Generate and Transfer Data with Rectified Flow},
  booktitle = {International Conference on Learning Representations},
  author = {Liu, Xingchao and Gong, Chengyue and Liu, Qiang},
  year = 2022
}

@article{ao2024overcoming,
  title={Overcoming the effect of muscle fatigue on gesture recognition based on {sEMG} via generative adversarial networks},
  author={Ao, Jinxin and Liang, Shili and Yan, Tao and Hou, Rui and Zheng, Zong and Ryu, JongSong},
  journal={Expert Systems with Applications},
  volume={238},
  pages={122304},
  year={2024},
  publisher={Elsevier}
}

@article{coelho2023novel,
  title={A novel {sEMG} data augmentation based on WGAN-GP},
  author={Coelho, Fabr{\'i}cio and Pinto, Milena F. and Melo, Aur{\'e}lio G. and Ramos, Gabryel S. and Marcato, Andr{\'e} L. M.},
  year={2023},
  journal={Computer Methods in Biomechanics and Biomedical Engineering},
  volume={26},
  number={9},
  pages={1008--1017},
  issn={1476-8259},
  doi={10.1080/10255842.2022.2102422},
  pmid={35862582}
}

@article{venugopal2024boosting,
  title={Boosting EEG and ECG Classification with Synthetic Biophysical Data Generated via Generative Adversarial Networks},
  author={Venugopal, Archana and Resende Faria, Diego},
  year={2024},
  journal={Applied Sciences},
  volume={14},
  number={23},
  pages={10818},
  publisher={Multidisciplinary Digital Publishing Institute},
  issn={2076-3417},
  doi={10.3390/app142310818},
  copyright={http://creativecommons.org/licenses/by/3.0/}
}

@article{dai2023improved,
  title={Improved Network and Training Scheme for Cross-Trial Surface Electromyography ({sEMG})-Based Gesture Recognition},
  author={Dai, Qingfeng and Wong, Yongkang and Kankanhali, Mohan and Li, Xiangdong and Geng, Weidong},
  journal={Bioengineering},
  volume={10},
  number={9},
  pages={1101},
  year={2023},
  doi={10.3390/bioengineering10091101}
}

@article{chen2025waveformer, 
    title = {WaveFormer: A Lightweight Transformer Model for {sEMG}-based Gesture Recognition}, 
    author = {Chen, Yanlong and Orlandi, Mattia and Rapa, Pierangelo Maria and Benatti, Simone and Benini, Luca and Li, Yawei}, 
    year = {2025},
    journal = {arXiv preprint arXiv:2506.11168},
    doi = {10.48550/arXiv.2506.11168}
}

@article{karnam2022emghandnet, 
    title = {EMGHandNet: A hybrid CNN and Bi-LSTM architecture for hand activity classification using surface {EMG} signals}, author = {Karnam, Naveen Kumar and Dubey, Shiv Ram and Turlapaty, Anish Chand and Gokaraju, Balakrishna},
    year = 2022, 
    journal = {Biocybernetics and Biomedical Engineering}, volume = {42}, 
    number = {1}, pages = {325--340}, issn = {0208-5216}, doi = {10.1016/j.bbe.2022.02.005} }

@inproceedings{kynkaanniemi2019improved,
  title={Improved Precision and Recall Metric for Assessing Generative Models},
  author={Kynk{\"a}{\"a}nniemi, Tuomas and Karras, Tero and Laine, Samuli and Lehtinen, Jaakko and Aila, Timo},
  booktitle={Advances in Neural Information Processing Systems},
  year={2019}
}

@inproceedings{naeem2020reliable,
  title={Reliable Fidelity and Diversity Metrics for Generative Models},
  author={Naeem, Muhammad Ferjad and Oh, Seong Joon and Uh, Youngjung and Choi, Yunjey and Yoo, Jaejun},
  booktitle={International Conference on Machine Learning},
  year={2020}
}

@article{xu2024chatemg,
  title={ChatEMG: Synthetic data generation to control a robotic hand orthosis for stroke},
  author={Xu, Jingxi and Wang, Runsheng and Shang, Siqi and Chen, Ava and Winterbottom, Lauren and Hsu, To-Liang and Chen, Wenxi and Ahmed, Khondoker and La Rotta, Pedro Leandro and Zhu, Xinyue and others},
  journal={IEEE Robotics and Automation Letters},
  year={2024},
  publisher={IEEE}
}

@inproceedings{sivakumar2024emg2qwerty,
  title={{emg2qwerty}: A large dataset with baselines for touch typing using surface electromyography},
  author={Sivakumar, Viswanath and Seely, Jeffrey and Du, Alan and Bittner, Sean and Berenzweig, Adam and Bolarinwa, Anuoluwapo and Gramfort, Alex and Mandel, Michael},
  booktitle={Advances in Neural Information Processing Systems},
  volume={37},
  pages={91373--91389},
  year={2024}
}

@article{atzori2014characterization,
  title={Characterization of a benchmark database for myoelectric movement classification},
  author={Atzori, Manfredo and Gijsberts, Arjan and Kuzborskij, Ilja and Elsig, Simone and Hager, Anne-Gabrielle Mittaz and Deriaz, Olivier and Castellini, Claudio and M{\"u}ller, Henning and Caputo, Barbara},
  journal={IEEE Transactions on Neural Systems and Rehabilitation Engineering},
  volume={23},
  number={1},
  pages={73--83},
  year={2014},
  publisher={IEEE}
}

@inproceedings{ho2020denoising,
  title={Denoising diffusion probabilistic models},
  author={Ho, Jonathan and Jain, Ajay and Abbeel, Pieter},
  booktitle={Advances in Neural Information Processing Systems},
  volume={33},
  pages={6840--6851},
  year={2020}
}

@article{prabhavathy2024hand,
  title={Hand gesture classification framework leveraging the entropy features from {sEMG} signals and VMD augmented multi-class SVM},
  author={Prabhavathy, T and Elumalai, Vinodh Kumar and Balaji, E},
  journal={Expert Systems with Applications},
  volume={238},
  pages={121972},
  year={2024},
  publisher={Elsevier}
}

@article{neifar2025deepa,
  title={Deep generative models for physiological signals: A systematic literature review},
  author={Neifar, Nour and Mdhaffar, Afef and {Ben-Hamadou}, Achraf and Jmaiel, Mohamed},
  year={2025},
  journal={Artificial Intelligence in Medicine},
  volume={165},
  pages={103127},
  issn={0933-3657},
  doi={10.1016/j.artmed.2025.103127}
}

@inproceedings{song2020denoising,
  title={Denoising Diffusion Implicit Models},
  booktitle={International Conference on Learning Representations},
  author={Song, Jiaming and Meng, Chenlin and Ermon, Stefano},
  year={2020}
}

\end{document}